\documentclass[11pt]{article}
\usepackage{graphicx,natbib}
\usepackage{subfigure}
\usepackage{epsfig}
\usepackage{epstopdf}
\usepackage[noend]{algpseudocode}
\usepackage{lineno}
\usepackage{algorithmicx,algorithm}

\usepackage{color,graphicx}
\usepackage{natbib,hyperref}
\usepackage{doi}

\usepackage{amssymb}
\usepackage{amsmath}
\usepackage{amsthm}

\newtheorem{myDef}{Definition}

\begin{document}

\title{Neighborhood Information-based Probabilistic Algorithm for Network Disintegration}

\author{Qian Li$^{a,*}$ and San-Yang Liu$^{a}$ and Xin-She Yang$^{b}$ \\
$^{a}$School of Mathematics and Statistics, \\
Xidian University, Xi'an, Shaanxi 710071, P.R. China \\
$^{b}$School of Science and Technology,\\
 Middlesex University, London NW4 4BT, UK \\
*Corresponding author}

\date{}

\maketitle 

\begin{abstract}
{\color{black}
Many real-world applications can be modelled as complex networks, and such networks include the Internet, epidemic disease networks, transport networks, power grids, protein-folding structures and others. Network integrity and robustness are important to ensure that crucial networks are protected and undesired harmful networks can be dismantled. Network structure and integrity can be controlled by a set of key nodes, and to find the optimal combination of nodes in a network to ensure network structure and integrity can be an NP-complete problem. Despite extensive studies, existing methods have many limitations and there are still many unresolved problems. This paper presents a probabilistic
approach based on neighborhood information and node importance, namely, neighborhood information-based probabilistic algorithm (NIPA). We also define
a new centrality-based importance measure (IM), which combines the contribution ratios of the neighbor nodes of each target node and two-hop node information. Our proposed NIPA has been tested for different network benchmarks and compared with three other methods: optimal attack strategy (OAS), high betweenness first (HBF) and high degree first (HDF). Experiments suggest that the proposed NIPA is most effective among all four methods. In general, NIPA can identify the
most crucial node combination with higher effectiveness, and the set of optimal key nodes found by our proposed NIPA is much smaller than that by heuristic centrality prediction. In addition, many previously neglected weakly connected nodes are identified, which become a crucial part of the newly identified optimal nodes. Thus, revised strategies for protection are recommended to ensure the safeguard of network integrity. Further key issues and future research topics are also discussed. \\

\hrule
\vspace{5pt}

\noindent {\bf Citation Details:} \\
Qian Li, San-Yang Liu, Xin-She Yang, Neighborhood information-based probabilistic algorithm for network disintegration,  {\it Expert Systems with Applications}, Volume 139, (2020), Article 112853.} \\
https://doi.org/10.1016/j.eswa.2019.112853

\hrule
\end{abstract}

\section{Introduction}
\label{S:1}

Network structures and characteristics appear naturally in many systems and applications. Examples are power grid networks, communication networks, transport networks, the Internet and others. Such networks form a crucial part of modern infrastructure, and the robustness and integrity of such networks can have a huge impact on the quality of life and society~\citep{buldyrev2010catastrophic,Chan2014Make,Watts1998Collectivedynamics,Shargel2003Optimization}. {\color{black}Improper management of such networks can be detrimental to society and economy.
The understanding of the complexity of network structures, stability of networks, and integrity of networks all require sophisticated theory and methods so as to estimate the safety and robustness of various networks in real-world applications. Therefore, the research on network robustness is both of theoretical interest and practical importance~\citep{Tanizawa2005Optimization,Chan2014Make,Shargel2003Optimization}. }

Existing studies have suggested that a small number of key nodes (or highly influential nodes) exist in most networks, and such nodes and their connectivity are crucially important to maintain the basic network structure and its performance~\citep{Adamic2000Power,Cohen2000Resilience,Watts2002A}.
On the one hand, we can protect the network by safeguarding key nodes and edges. On the other hand, the isolation of certain key nodes and edges (such as diseases and fires), the network can be potentially made to collapse \citep{Holme2002Attack}. Therefore, network disintegration (or network dismantling) becomes an important research hotspot, and  many studies have devoted to this area~\citep{braunstein2016network,yang2017vulnerability,radicchi2015predicting}.

{\color{black}The essence of network disintegration problem is to find the best (often the smallest) set of key nodes that are crucial to the integrity and robustness of the networks. Many infrastructures and societal functionalities are dependent on such networks, including the smooth supply of energy, water, food and resources, control of diseases, protection of environment and many others~\citep{jain2019discover}.
The understanding of such networks and the insights gained can be very useful to many applications such as business planning, smart cities and smart grids, protection of crucial networks such as the Internet and power grids to avoid any large-scale failures. All these are relevant to expert and intelligent systems.
Thus, the studies of network integrity and robustness can have some profound implications on social and economical activities.}

Most existing studies heuristically ranked the importance of nodes of networks using difference measures, usually based on graph theory~\citep{Watts2002A,Borgatti2013Centrality,Freeman1977A}. In the current literature,  centrality based measures seem to be the most widely used \citep{Borgatti2013Centrality,Freeman1977A}. Measures such as the degree~\citep{Pastorsatorras2001Epidemic,Cohen2001Breakdown}, betweenness centrality~\citep{borgatti2005centrality,bai2017effective}, k-core~\citep{Kitsak2010Identification}, PageRank~\citep{Brin2008The}, eigenvector centrality~\citep{Straffin1980Linear} and closeness centrality~\citep{Bavelas1950Communication} have all been applied to the disintegration problems.

{\color{black}
However, the heuristic rank strategy may not be a good measure for network disintegration problems because it essentially combines some isolated nodes with high centrality into a set,  without counting the interactions among different nodes. As Braunstein pointed out that this is an essentially collective problem, the optimal dismantling set cannot be considered as a collection of well-performing nodes~\citep{braunstein2016network}.} Recent studies also indicated that a certain small specific set of nodes could determine the main network structure, and a large number of previously neglected weakly-connected vulnerable nodes might be crucial to the overall network cascades~\citep{Yang2017Small}. An important issue is that the existing centrality-based methods struggle to find such key, optimal combination of nodes because global optimization was not used in this case~\citep{morone2015influence}, though some estimation of robustness can be carried out~\citep{Wandelt2018QRE}.

{\color{black}
The problem for finding the optimal set of key nodes to protect or destroy is a non-deterministic polynomial-time (NP) hard problem. In fact, it is NP-complete~\citep{morone2015influence,Johner2009Optimal}. Obviously, such hard problems do not permit exhaustive search for optimal solutions, and thus heuristic and approximate methods are often used to find a good set of feasible (sometime, optimal) solutions. This can be achieved either by using various problem-specific heuristic strategies or by using black-box type metaheuristic optimization methods in combination with traditional methods~\citep{Guturu2008An}. Despite these studies, there are still many challenging issues to be resolved.}

Motivated by the above challenges and issues, we propose a novel neighborhood information-based probabilistic algorithm (NIPA) for network disintegration, which considers not only the two-hop neighbor nodes but also the new quantitative measures for centrality. Thus, the main contributions of this work can be summarized as follows:
\begin{enumerate}
  \item  A novel centrality-based importance measure (IM) is proposed in this paper, which essentially measures the importance of target nodes in networks under attack. The two-hop node information of the target node is combined with the contribution ratio, which can overcome the limitation of traditional degree measures in terms of a single piece of nodal information.

  \item Group effects have been considered in our approach where a reservation mechanism with the combined influence for the set of best attack nodes is proposed for candidate solutions.

  \item Based on a reservation mechanism and attack probability related to IM, a heuristic probabilistic algorithm is proposed, which uses a probabilistic preferential selection for node attack in consideration of dynamic network structures.
\end{enumerate}

Therefore, the paper is organized as follows. {\color{black} Section 2 reviews the related recent developments, and Section 3 provides some background concerning the optimization model for complex network disintegration, relevant algorithms and estimation of measures. Section 4 describes the proposed neighborhood information-based probabilistic algorithm (NIPA), including  the calculation of the attack probability, update strategy, and the realization of reservation mechanism. Section 5 summarizes the experiments on various network models, together with the comparison with three other methods. Then, Section 6 discusses parameters analysis and the complexity of the proposed algorithm. Finally, Section 7 concludes the paper with some discussions for future research. }

{\color{black}\section{Recent Developments}
\label{S:2}

Network disintegration is important to many applications, including protecting and safeguarding key networks such as power grids, transport networks and the communications network, and dismantling undesired networks such as diseases networks. Thus, it is no surprise that the studies of network integrity and robustness have become an active research topic. Various studies used different approach attempting to obtain optimal node sets for attacking and protecting a particular network.

In the current literature, centrality-based measures seem to be most widely used~\citep{Borgatti2013Centrality,Chan2014Make}, which has rigorous mathematical foundation based on graph theory. The so-called high degree first (HDF) method is a classic approach, based on the degree centrality of nodes, and the attack strategy is to first attack those nodes with high degrees. This seems reasonable because, in most cases,  the destruction of a node with multiple edges can have a significant impact on the network. The advantage of HDF is its low complexity. However, the degree of a node is a local property, not relative to the entire network~\citep{Pastorsatorras2001Epidemic,Cohen2001Breakdown}. In other words, only one hop neighborhood information is considered in this approach. Consequently, it can be expected that the algorithm is not very effective.

The high betweenness first (HBF) approach is another classic method for network disintegration. Here, the betweenness centrality refers to the percentage of the shortest paths between any two nodes in the network passing through the certain node~\citep{rabade2014survey}. The high betweenness nodes are equivalent to   bridges connecting nodes. Compared with HDF, HBF has a higher complexity, but it uses some global property.
However, as more nodes are attacked and removed, the network structure changes, leading to potentially very different distributions of degrees and betweenness measures from those for the initial network.

Recently, researchers started to explore other measures of network structures, with the emergence of some new methods. For example, Anggraini et al. proposed a two-step method~\citep{Anggraini2015Network}, where the first step was to detect the communities in a given network and then to delete the links between them to obtain isolated communities. Its second step was to eliminate the key nodes in each community. This could be effective for large-scale social networks because the network structure could be greatly simplified after dividing into different communities and thus it became easier to detect the key nodes in the communities. However, the main limitation is that this method is only suitable for networks that are easily partitioned, and community partition itself is a more difficult problem, which has not been completely solved. Consequently, this method can be inefficient for many real-world networks.

A different approach was proposed by Braunstein et al., based on statistical mechanics and a three-stage minimum sum algorithm for efficient decomposition networks~\citep{braunstein2016network}. This algorithm
consisted of three stages: a) decycling by a message-passing variant, b) breaking large trees/components into small ones, and c) re-inserting some nodes to close cycles. Though this method can obtain good results, it is more suitable for networks with a light-tailed distribution of degrees. In addition, the
complexity of this algorithm is high.

Furthermore, Tan et al.~proposed a method for finding key nodes by using link prediction~\citep{tan2016efficient}, which was suitable for networks with missing information because link prediction was used to recover some links. However, this method may over-predict the lost links, and thus limits the use and performance of this approach. Alternatively, network problems may be mapped onto random networks network with percolation actions, which can model the epidemic process as diffusion over networks~\citep{morone2015influence}.

On the other hand, heuristic optimization algorithms have shown to be a good alternative to solve NP-hard problems when exact methods are impractical for large-scale problems. Many metaheuristic algorithms have been successfully used to solve NP-hard problems such as the travelling salesman problems and vehicle routing problems. These new optimization algorithms include ant colony optimization (ACO)\citep{Dorigo2002Ant}, particle swarm optimization (PSO) \citep{Huang2004Particle,Shia2007Particle}, genetic algorithm (GA) \citep{Chatterjee1996Genetic,Marinakis2007A}, firefly algorithm (FA)~\citep{Osaba2017},
bat algorithm (BA)~\citep{Osaba2016} and other  swarm intelligence based algorithms~\citep{Meng2016Population,Li2015Colored}. These algorithms can obtain surprisingly good solutions with sufficient accuracy.

However, metaheurisitc algorithms have not been well studied in the context of network robustness or integrity, though there were some preliminary studies in this area. For example, Deng et al. presented an optimized attack strategy model for complex networks and used the tabu list for solving network disintegration problems so as to identify the optimal attack combination~\citep{Deng2016Optimal}. Their approach transformed the network disintegration problem into a binary integer programming problem.
Compared with the classical methods, numerical simulations showed that their attack strategy could improve the effect of network disintegration. However, greedy approaches were used in their approach, which means that there is no guarantee that the optimal attack strategy can be found. In addition, this approach did not consider the nodal interactions and information of network structures, which can significantly limit its performance.

Our new approach will use a new measure by considering the interactions of nodes and neighborhood information of the networks. A neighborhood information based probabilistic algorithm (NIPA) is proposed for finding the optimal attack strategy in terms of key nodes and node combinations.}

\section{Background}
\label{S:2}

\subsection{Definitions}
Let graph $G = (V,E)$ represent an undirected network with $N = \left| V \right|$ nodes and $M = \left| E \right|$ edges. Let $A = {({a_{ij}})_{N \times N}}$ be the adjacency matrix of the graph $G$, where ${a_{ij}}~{\rm{ = }}~{a_{ji}} = 1$ if nodes ${{\rm{v}}_i}$ and ${{\rm{v}}_j}$ are adjacent.

For a node (say, node $i$, connect to node $j$ in the largest connected cluster) on a network with a connectivity probability distribution $P(k)$, a percolation transition can occur when this node ($i$) is also connected to at least one other node, which means that the average degree of nodes must be at least 2. Otherwise, the largest cluster may become fragmented~\citep{Cohen2000Resilience}. This means that there is a critical disintegration threshold:
\begin{equation}
\left\langle {{k_i}\left| {i \leftrightarrow j} \right.} \right\rangle  = \sum\limits_{{k_i}} {{k_i}P({k_i}\left| {i \leftrightarrow j} \right.)}  = 2,
\label{Eq-100}
\end{equation}
where $k_i$ is the degree of connection of node $i$.

Using the Bayesian rule for conditional probabilities, we have
\begin{equation}
  P({k_i}~|{\rm{ }}i \leftrightarrow j) = \frac{{P({k_i}~|{\rm{ }}i \leftrightarrow j)}}{{P(i \leftrightarrow j)}} = \frac{P(i \leftrightarrow j|{\rm{ }}{~k_i})P({k_i})}{P(i \leftrightarrow j)},
\end{equation}
where $ P({k_i}~|{\rm{ }}i \leftrightarrow j)$ corresponds to the probability that the degree of node $i$ is equal to $k_i$ when there is a connection to node $j$.

For connected networks (ignoring loops), we have $P(i \leftrightarrow j) = \left\langle k \right\rangle /(N - 1)$ and $P(i \leftrightarrow j\left| {{k_i}} \right.) = {k_i}/(N - 1)$, where $N$ is the total number of nodes in the network.
Thus, Eq.~(\ref{Eq-100}) becomes
\begin{equation}
\kappa \equiv \frac{<\! k^2\!>}{<\! k \!>} = 2.
\end{equation}
In the framework of percolation theory,  when the critical condition ${\bf{\kappa}}=2$ occurs, networks can be considered as completely disintegrated~\cite{Cohen2001Breakdown}.  Thus, $\kappa > 2$ means that the network is not disintegrated. To measure the extent of the network integrity, the classical measure $S(Q)$ is often used, which is essentially the fraction of nodes in the largest connected cluster after removing $Q$ nodes ($Q=qN$ where $q$ is the attack ratio). In essence, $Q$ can be considered as a measure of the attack intensity whose value varies from $0$ (no attack or removal) to $N$ (all nodes are attacked or removed). In addition, in order to measure the effectiveness of an attack strategy, a minimal fraction ${q_c}$ of nodes is used when the network under attack is considered as completely collapsed \citep{Valente2004Two,Albert2000Error}.
However, this measure does not consider the situations where a network may suffer a big damage without completely collapsing.

In order to remedy this drawback, some researchers used a measure that considers the size of the largest connected cluster during all possible malicious attacks~\citep{Schneider2011Mitigation,Cohen2000Resilience,Cohen2001Breakdown,Herrmann2011Onion},
which intends to take into account the extent of damage for nodal removal. Thus, a unique robustness measure $R$ can be defined in the following manner~\citep{Schneider2011Mitigation}:
\begin{equation}
R = \frac{1}{{N + 1}}\sum\limits_{Q = 0}^N {S(Q)}.
\end{equation}
Its range of values is between 0 and 0.5, which loosely corresponds to an original network with isolated nodes ($R$ is close to 0) to the most robust, fully connected networks ($R=0.5$).

\subsection{Optimization Formulation}
The status of a node $i$ can be denoted by a binary variable: either for existence ($x_i=1$) or removed/attacked ($x_i=0$). Thus, all the $N$ nodes form
a fixed-length binary string $({x_1},{x_2}, \cdots {x_N})$ where $ \forall x_i\in\{0,1\}$.
With this representation, the problem of network disintegration can be transformed the following a binary integer programming problem of fixed-length binary decision variables in $N$ dimensions, the goal is to find a qualified binary string to minimize $S(Q)$~\citep{Deng2016Optimal}:
\begin{equation}
\centering
\begin{array}{l}
\textrm{Minimize } S(Q), \\
\textrm{subject to }
\left\{ {\begin{array}{*{20}{l}}
{Q = N - \sum\limits_{i = 1}^N {{x_i}}, }   \\
x_i \in \{0, 1\}, \quad (i=1,2,...,N).
\end{array}} \right.
\end{array}
\end{equation}

 Specially, this binary integer programming problem is still very hard to solve. However, it is possible to use some heuristic algorithms to solve such problems. The population-based incremental learning algorithm is one example~\citep{Larraanaga2001Estimation}, where they used a heuristic approach based on probability to solve binary discrete optimization problems. However, it was not used for
solving network problems.

\subsection{Optimal attack strategy (OAS)}

The optimal attack strategy (OAS) \citep{Deng2016Optimal} has been used to solve the above type of network optimization problems, which can be considered as a heuristic evolutionary algorithm applied to study network robustness. In this approach, a population of candidate solutions are represented in terms of a set of fixed length binary strings of length $N$ (where $N$ is the total number of nodes). Each bit of a string only takes two values $1$ or $0$, where $1$ represents the existence of the node in the network, while $0$ means that the corresponding node is attacked and removed.
Initially, a vector ${(1, 1, \cdots, 1)_{1 \times N}}$ is generated, then $Q$ nodes are removed randomly to obtain the first initial solution as a starting solution. By swapping the states of two nodes randomly (by interchanging 0 and 1), a set of ${n_p}$ population individuals are generated. The candidate solution with the best $S(Q)$ is passed onto the next generation without swapping. A tabu list for nodes was used in this approach.

The OAS has been compared with other four methods \citep{Deng2016Optimal}, which showed that the OAS could improve the effect of network disintegration. However,
OAS is quite computationally extensive because it does not have a good iterative and update mechanism; it is somehow equivalent to an exhaustive search for the best solution by all possible scenarios. Thus, the algorithm performance is limited.

In order to overcome these shortcomings, we propose a heuristic optimization algorithm, called neighborhood information-based probabilistic algorithm(NIPA). The main difference from the OAS is that the proposed NIPA has a more effective iterative mechanism: the probability of target nodes to be attacked is calculated at each iteration. Briefly speaking, the probability vector is obtained through a new centrality measure,  importance measure (IM), and next generation of target attack nodes is guided by probability. Another main difference from OAS is that the effect of grouping is also considered in our algorithm, which selects a better combination of nodes so as to effectively attack the network in the next generation, which can be considered as an attack strategy through the principle of probability priority.
Loosely speaking, a node, which lacks superiority in probability, will be replaced, and the probability vector is proposed via the dynamic change of network structure. With these key modifications, NIPA can have significant advantages in terms of attack effectiveness, compared with the classical methods.
This will be demonstrated in the simulations later in this paper.
Now let us introduce some details of the algorithm.

\section{Neighborhood information-based probabilistic algorithm}
\label{S:3}

Finding the best attack combination of nodes in a network can be a computationally extensive task. In this paper, our proposed NIPA can be considered as an evolutionary algorithm, which uses neighborhood information of node distributions, dynamic probability, a reservation mechanism, and update strategy. Their details will be outlined below.

\subsection{Initialization}

The initial attack should not be random. Rather, it should use some prior information, based on some knowledge of the network structure. Ideally, the initial solutions should be not only easy to obtain, but also effective. In addition, it should also potentially reduce the computational complexity, though this cannot be achieved easily. However, taking account of the connectivity of nodes, the initialization can use some information, based on some conventional centrality measures.

In the current literature, centrality measures such nodal degrees have been frequently used to evaluate the importance of nodes  in the network~\citep{Everett2005Extending}. In this paper, initialization is carried out, according to the degrees of nodes. This is achieved by sorting the degrees of nodes in the descending order and then selecting $Q$ nodes with the larger degrees as the initial attack nodes. The initial solution ${X^{now}}$ is an $N$-dimensional binary string. Those selected attack nodes correspond to locations/bits with $0$s, while the remaining, unattacked, nodes are 1s. It is worth pointing out that other centrality measures, such as betweenness and closeness, can also be used.

It has been observed from our simulations that the sole reliance on the degrees of nodes to attack the network is not sufficient. As we will see later, some neighborhood information is needed to increase the attack effectiveness.

\subsection{Definition of a centrality measure: important measure}

Conventional centrality measures only focus on a particular property of nodes.
Consequently, many existing studies concerning attack strategies on complex networks focused on certain properties of the nodes such as the degrees of nodes \citep{Holme2002Attack}. For example, many strategies attacked nodes with most attached edges with a hope to maximize the damage. Despite such rationality, the effectiveness of such attacks is not as high as it is expected. In many cases, just use of the degree information is not adequate. Thus, there is some information missing from such strategies.

\begin{figure}[ht]
\centering
\includegraphics[width=6cm]{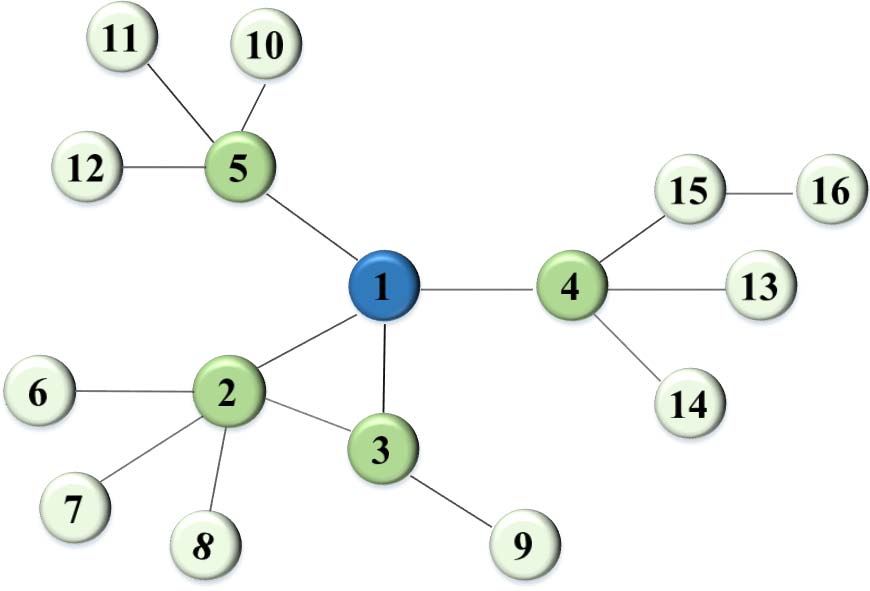}
\caption{Node 2 is the highest degree, but it is not the one that should be attacked first. It is obvious that an attack on Node 1 can be more disruptive to the network.}
\end{figure}

For example, as shown in Figure 1. The degree of Node 2 is 5, while the degrees of Nodes 1, 4 and 5 are all 4, the degree of Node 3 is 3. If the principle of the nodal degree  attack algorithm is used, Node 2 should be attacked first. But it is obvious that an attack on Node 1 can fragment the network into three isolated clusters, which is far more damaging to the network than the effect of attacking Node 2.

Therefore, this highlights an issue that the degree-based strategy is not sufficient.
We urgently need a new centrality measure to quantify the importance of nodes.
Since our task is to minimize the objective $S(Q)$, the nodes to be attacked should be those nodes whose removal can be detrimental to the network, and also may lead to the separation of more nodes from the largest connected cluster.

In order to capture such observations, a novel centrality measure is introduced here: the importance measure (IM) for network disintegration is defined as follows:
\begin{myDef}
The contribution ratio of a neighbor node is defined as the probability, equal to the reciprocal of the number of target attack nodes that are directly connected to this neighbor node. That is
\begin{equation}
{C_{{j_t}}} = \frac{1}{{\left| v \right|}}, \quad v = \{i \left| \; {a_{{j_t},i}} \right. = 1, \;\; i = 1,2, \cdots Q\} ,
\end{equation}
\end{myDef}
\noindent where $j$ ($j = 1,2, \cdots Q$) corresponds to the $j$-th attack node, and ${j_t}$ indicates the $t$-th neighbor node of the $j$-th attack node. The set $v$ represents the set of attack nodes (each node may connect more than one attack node) that are directly connected to ${j_t}$-th neighbor node. Here, $\left|  \cdot  \right|$ represents the number or cardinality of elements in the set $v$.

\begin{myDef}
The importance measure (IM) of a target attack node is defined as the sum of the product of the degree of each neighbor node and its contribution ratio. Such neighbor nodes are not included in the largest connected cluster of the network after the target attack node is deleted.
\begin{equation}
\begin{array}{l}
{{IM_j} = \sum\limits_{{j_t} = 1}^T {{C_{{j_t}}}} {k_{{j_t}}}}, \quad {j_t} \in \Theta  \cap \Omega_j, \\
\Theta  = \{ 1,2, \cdots ,N\}, \quad \Omega_j  = \left\{  j_t \left|  {\tilde a_{j,{j_t}}} = {\tilde a_{{j_t},j}} = 1,
  \; \tilde a \in \tilde A \right. \right\},
\end{array}
\end{equation}
\end{myDef}
\noindent where ${k_{{j_t}}}$ means the degree of the $t$-th neighbor of the $j$-th attack node. Here, $\Omega$ represents the set of the neighborhood nodes of the $j$-th attack node, and $T$ is the number of $\{ {j_t}\}$. Here, $\tilde A$ is an adjacent matrix,
which does not contain nodes in the largest connected
cluster after the target attack node is removed.

In essence, the IM measures not only the information of the target node, but also the information about the two-hop neighbor nodes. The next example illustrates this clearly. As seen in Figure 1, Node 1 and Node 2 are target nodes.
After attacking Nodes 1 and 2, the largest connected cluster of the remaining network contains Nodes 4, 13, 14, 15 and 16. After removing these 5 nodes, Node 1 has 3 neighbor nodes, which are Nodes 2, 3, and 5, respectively. In addition, Nodes 2 and 5 are simply connected to Node 1, their contribution ratio is 1. But Node 3 is not only the neighbor of Node 1, but it is also the neighbor of Node 2, which means that the contribution ratio of Node 3 to Node 1 is 1/2. As the degrees of Nodes 2, 3, and 5 are 5, 3,  and 4, respectively. The IM of Node 1 is obtained by summing the products of the contribution ratios and their corresponding degrees of the corresponding neighbor nodes. That is
\begin{equation}
{IM_1} = 1 \times 5 + \frac{1}{2} \times 3  + 1 \times 4 =10.5.
\end{equation}

Similarly, for Node 2, there are five neighbor nodes, which are Nodes 1,  3,  6,  7, and 8. The corresponding contribution ratios are 1, 1/2, 1, 1, and 1, respectively. Thus, its IM is
\begin{equation}
{IM_2}  = 1 \times 4 + \frac{1}{2} \times 3 + 1 \times 1 + 1 \times 1 + 1 \times 1 = 8.5,
\end{equation}
which shows that Node 2 is less important than Node 1. Thus, an attack on Node 1 should be prioritized.

It should be noted that the above calculation of the IM also considers the interactions of two attack nodes. For example, if two adjacent nodes are both attacked,
they are regarded as part of the neighborhood nodes. Thus, the above $\tilde A$ varies at every generation because the target nodes of each generation can be different. To illustrate how a target node is selected in our algorithm,
an attack probability, reservation mechanism and update strategy will be introduced in the next subsections.

\subsection{Attack probability}

The attack probability of a node is used to indicate the extent of damage to the network after that node is attacked. However, the calculation of individual probability of each node in a large network can be very time-consuming and unnecessary. A main advantage of our algorithm is that it needs to only compute the probabilities of target attack nodes (the bits with values of $0$ in a binary string) in the current generation at a time.

Based on the above definition of IM, we can calculate the normalized attack probability vector $P$ at each iteration as follows:
\begin{equation}
{P_j} = \frac{{{IM_j}}}{N},\quad  j = 1,2, \cdots Q \label{IM-eq-100}
\end{equation}

There are potentially many different ways of calculating probabilities, our simulations show that there is no significant difference among different methods. The present formulation is used for simplicity as well as numerical experiments.

\begin{algorithm}[!h]
    \caption{Calculate the attack probability}
    \begin{algorithmic}[1]
    \State  {\bf{Input}}: the adjacency matrix $A$, the attack intensity $Q$, the initial binary string ${X^{now}}$ (with $Q$ bits being 0, and $N-Q$ bits being $1$), the node $i$ degree $k_i$, and the number $N$ of nodes.
    \State  $\Phi \leftarrow node_{i}$: Get attack nodes index set $\Phi$
    \State $A_{temp} \leftarrow A$
    \For{$i=1,2,...,Q$}
    \State $A_{temp}(\Phi(i),:) \leftarrow \bf{0}_{1 \times N}$
    \State $A_{temp}(:,\Phi(i)) \leftarrow \bf{0}_{N \times 1}$
    \EndFor
    \State {\bf end for}
    \State Get the node set $\Im$ of the largest connected cluster from $A_{temp}$
    \For{$t=1,2,...,Q$}
    \State $\tilde A(\Im(t),:) \leftarrow \bf{0}_{1 \times N}$
    \State $\tilde A(:,\Im(t)) \leftarrow \bf{0}_{N \times 1}$
    \EndFor
    \State {\bf end for}
    \State $\tilde A \leftarrow A$
    \For{$l=1,2,\cdots,Q$}
    \State $\Omega(\Phi(l)) \leftarrow$ the neighbor nodes of $l$th element of $\Phi$
    \EndFor
    \State {\bf end for}
    \State Calculate the frequency of each node in $\Omega$
    \For {$j=1,2,\cdots,Q$}
    \If{$\Omega(\Phi(j))==0$}
    \State ${P(\Phi(j))=0}$
    \Else
    \State $P(\Phi(j))$ by Eq.(\ref{IM-eq-100})
    \EndIf
    \State ${\bf{end~if}}$
    \EndFor
    \State \bf{end for}
    \State $\mathbf{return}~{P}$ as a vector.
    \end{algorithmic}
\end{algorithm}

{\color{black}
The attack probability of each attack node is calculated, according to the steps outlined in Algorithm 1. The greater the probability of a node, the more nodes (not attack nodes) are to be separated from the largest connected cluster after attacking that node. In order words, the probability is the probability of a bit being $0$ in the binary string representations.
}

\subsection{Reservation mechanism}

From the graph theory for complex networks, we know that network integrity can largely depend on a specific (often small) subset of nodes, namely structural nodes. In many cases, such nodes are not necessarily always highly connected. Existing studies have indicated that some low-centrality nodes can play an important role in network integrity. Many previously neglected weakly connected nodes can surprisingly be crucial in the set of attack nodes~\citep{morone2015influence}. However, most existing studies do not consider the function or effect of the combination of such low-centrality nodes, but such nodes can be very crucial as shown in the following example.

\begin{figure}[th]
\centering\includegraphics[width=7cm]{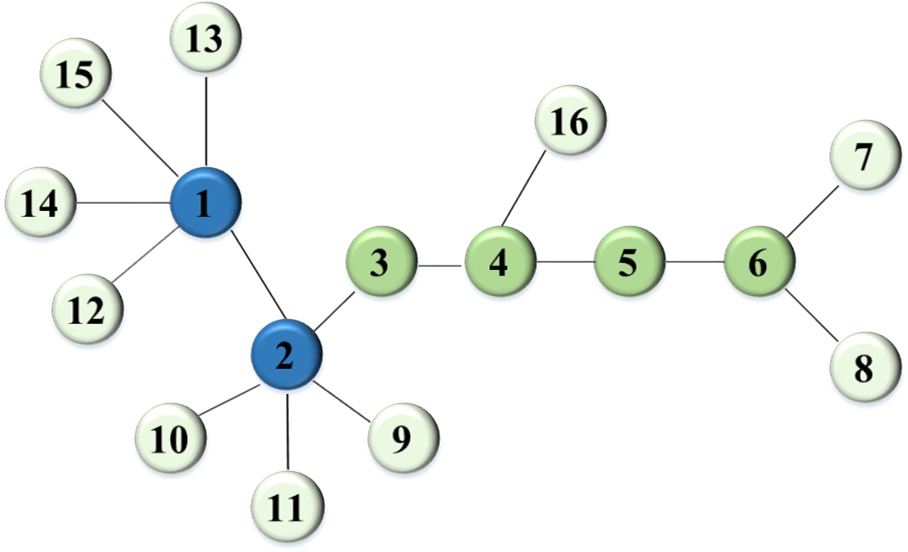}
\caption{Two nodes with a high degree (Node 1 and Node 2) are not a good combination for attack. Nodes 1 and 4 are a better combination for a detrimental attack.}
\end{figure}

As is shown in Figure 2, if an attack strategy is determined by the degrees of nodes, then Node 1 and Node 2 should be attacked. But the fraction of nodes in the largest connected cluster after removing those two nodes is 7/16. It is easy to see that it is more damaging to the network if Node 1 and Node 4 are attacked, which leads to a fraction (5/16) of nodes in the largest connected cluster.

Therefore, the attack on a combination of lower-degree nodes may be more effective than that on the nodes of higher degrees. In order to use such crucial information of node combination in our proposed algorithm, we should select the best solution(s) among all the candidate solutions at each generation. This solution is effective, probably because of the combination of some specific nodes, and the inheritance of such best combinations to the next generation. This acts as a reservation mechanism, in a similar fashion as elitism in genetic algorithms and many other algorithms \citep{Yang2014}. In addition, it becomes computationally extensive if certain quantities such as adjacent matrices are to be calculated at each iteration.
It may be time-saving if we use the sparsity of the adjacent matrix and only
update the part around the attacked node because the entries for unattacked nodes
will largely remain unchanged. It is worth pointing out that the reservation list
is not a tabu list. Rather, it is a form of elitism as a time-saving strategy using the sparsity characteristics of adjacent matrices to ensure that the best solutions
can be passed onto the next generation so as to improve the rate of convergence.

{\color{black}
Thus, the reservation mechanism has been implemented in this paper. At each iteration, a certain number of attack nodes (in percentage) are kept unchanged. That is, $\Lambda  = \alpha Q$ where  $\alpha  \in (0,1)$, though we have typically used $\alpha=0.3$ in our implementation, based on a basic parametric study. In the next iteration, these reserved nodes remain unchanged and only the other $Q - \Lambda $ nodes are updated accordingly.}

\subsection{Update strategy}

Since $\Lambda $ nodes are kept unchanged in the reserve list, there are only $Q - \Lambda$ attack nodes to be updated.
To be more specific, in the binary string, a node with a status of 1 is randomly selected and its state is changed from 1 (exist) to 0 (removed). Then, the state of another attack node with a state of 0, which is not a reservation node,  is changed from 0 to 1 (exist), as shown in Fig.~\ref{Fig-3}.
This acts somehow as a mutation mechanism. In essence, this update can be considered as a local random walk \citep{Yang2014}.

\begin{figure}[th]
\label{Fig-3}
\centering\includegraphics[width=10cm]{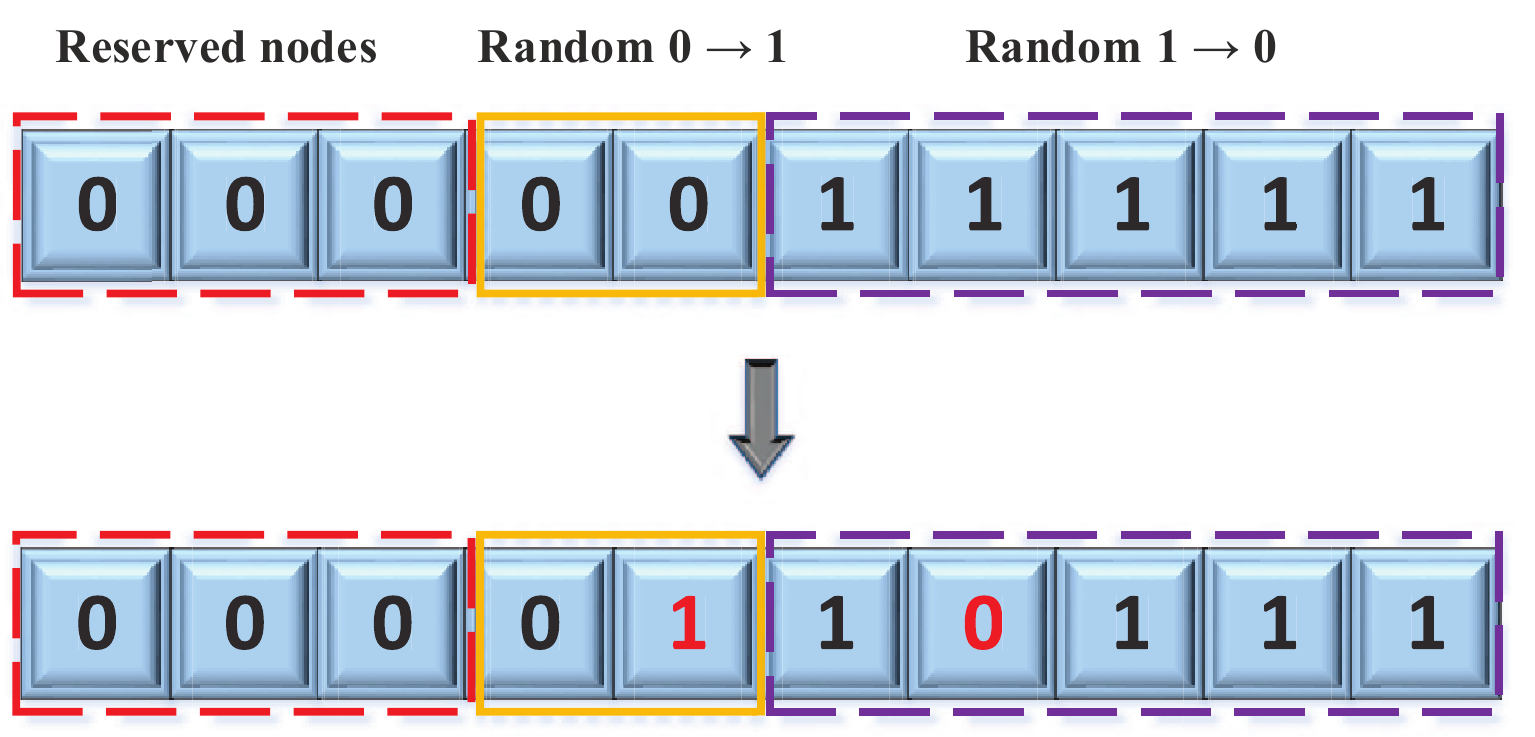}
\caption{Mutation scheme as an update strategy where a bit of 0 in the string represents the attack node. If the first 3 bits represent the reserved nodes, we can choose one of the other two nodes to convert from 0 to 1. At the same time, we need to randomly select another node with bit 1 and change it from 1 to 0.
\label{Fig-3} }
\end{figure}

\subsection{The main steps of our algorithm}

The above descriptions of the proposed NIPA can be summarized as the following seven steps:

\begin{itemize}
\item Step 1:   Initialization of the parameters: attack strength $Q$, total number of nodes $N$, population size $n_p$. Initialize the population as ${C_{best}}$, and calculate $S{(Q)_{best}}$.
\item Step 2:   Termination condition: If the attack intensity $Q$ is known, iterations stop when the number of iterations reaches
    $T= $iteration$_{\max}$. In order to compare the performance of the algorithm with other algorithms and to ensure consistency, iterations stop if $Q > N$(or $q > 1$) or $\kappa  \le 2$.
\item Step 3:   Calculate the attack probability vector: Calculate the largest connected cluster after attacking $Q$ nodes, then remove the nodes contained in the largest connected cluster from the original network. Calculate the attack probability by formula (\ref{IM-eq-100}).
\item Step 4:   Reservation list: Sort the attack probability of $Q$ nodes in descending order, and retain $\Lambda$ nodes with higher probability.
\item Step 5:   Update Strategy: For non-reserved attack nodes, select one node randomly to exchange with the unattacked node.
\item Step 6:   Calculate $S(Q)$, select the best value in the population, denoted by $S{(Q)_{now}}$, and the corresponding individual recorded as ${C_{now}}$.
\item Step 7:   Determine if $S{(Q)_{now}}$ is better than $S{(Q)_{best}}$. If it is true, assign the value of $S{(Q)_{now}}$, ${C_{now}}$ to $S{(Q)_{best}}$, ${C_{best}}$. Go to Step 2.
\end{itemize}

\begin{figure}[th]
\centering\includegraphics[width=10cm]{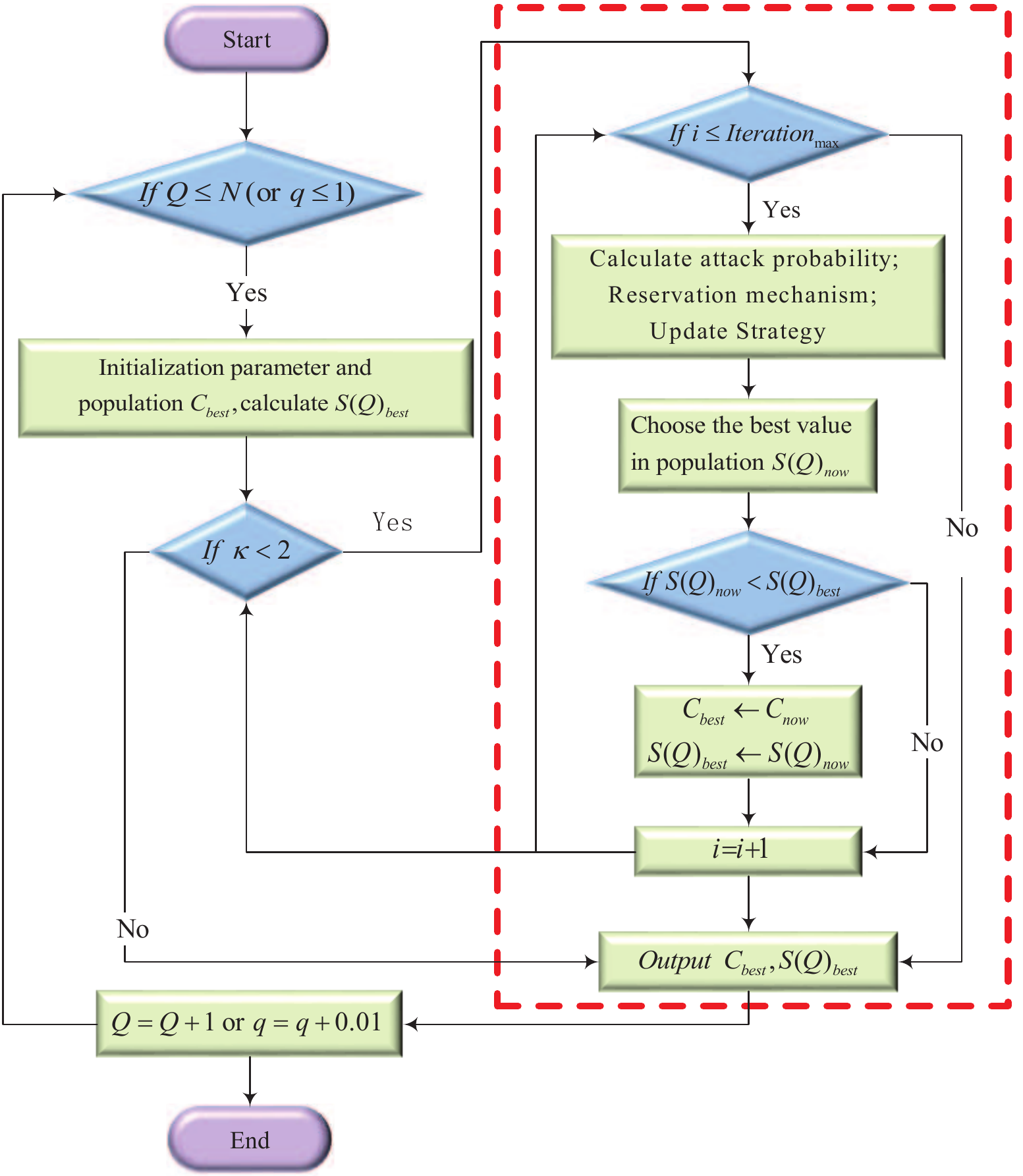}
\caption{The schematic representation of NIPA. For a given attack intensity, the steps
for initialization and execution are highlighted inside the dashed box. \label{Diagram-loop} }
\end{figure}

The steps and main loops can be visualized in Fig.~\ref{Diagram-loop}. In the rest of the paper, we will use the proposed approach to carry out a series of numerical experiments.

\section{Numerical Experiments}
\label{S:4}

In order to validate the performance of the proposed NIPA, a series of experiments have been carried out using the standard karate instances
and different network models.
First, we apply our NIPA to the karate club network, which is a friendship network with 34 members of a karate club at a US university, as first outlined by W. Zachary in 1977 \citep{Zachary1977An}. A comparison has been made between the proposed algorithm
and three other algorithms: optimal attack strategy (OAS), high degree first (HDF), and high betweenness first (HBF). The OAS \citep{Deng2016Optimal} is a recently proposed heuristic approach based on the well-known tabu search, while HDF \citep{Wang2007A} and HBF are conventional algorithms for network disintegration, using certain information related to the degree and betweenness centrality, respectively. In addition, further comparison and evaluations are also carried out on three different network models: the Barab\'asi-Albert (BA) scale-free network~\citep{barabasi1999emergence,Albert2001Statistical,Adamic2000Power}, the Erd\H{o}s-R\'enyi (ER) random network model \citep{Erd2012On}, and the Watts-Strogatz (WS) small-world network \citep{Watts1998Collectivedynamics,Yang2001}.

All the algorithms have been implemented using Matlab version 2017b on a computer with a CPU i7-8700 processor, 8GB RAM, running Windows 10.  In the rest of this paper, we will discuss each numerical experiment and present the results in detail.

\subsection{The karate club network}

The karate network is a small-scale network, but it can be considered as a microcosm of realistic interpersonal or social networks. In this network, socially active people with many friends have more connections/edges, while solitary people may have only a single link connecting to the network. The main aim is to see who can sustain the relationship in the network under various attacks.
We will use our NIPA and three other algorithms to identify the key people (nodes) in the club (network), and the removal of such nodes will potentially disrupt the network.

\begin{figure}[ht]
\centering
\includegraphics[width=4.5in]{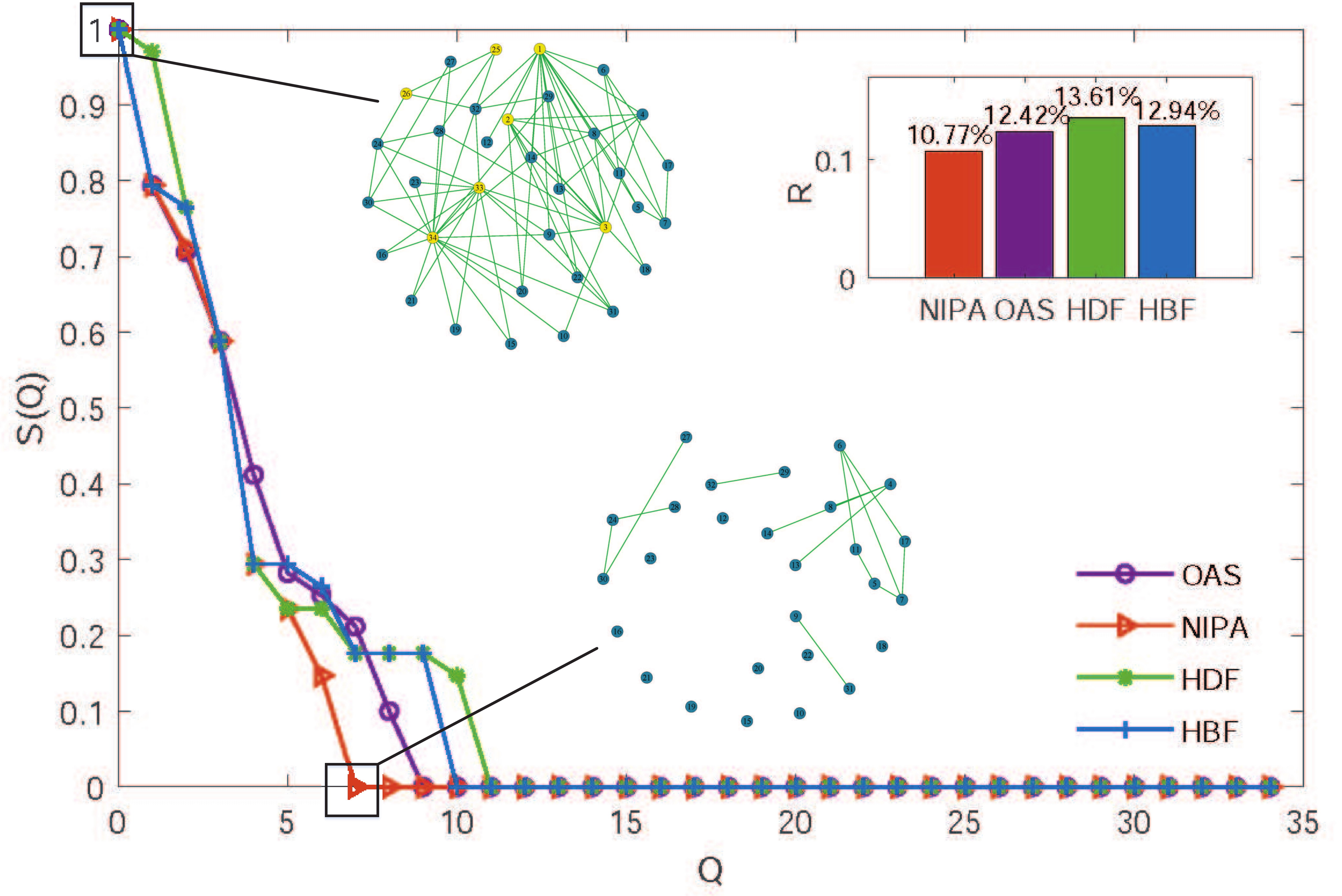}
\caption{Comparison of four attack methods on the karate network. \label{KN-fig} }
\end{figure}

The population size is 100, and the maximum number of iterations is set to 100. In addition, the percentage of reservation $\alpha$ is set to 0.3,  and set the tabu list to 10 in OAS. The results of the numerical simulation using four methods are summarized in Fig.~\ref{KN-fig}.

As we can see from the inset of Fig.~\ref{KN-fig},
the $R$ value of our NIPA is much smaller than the other three methods, which means that our proposed algorithm can use a relatively smaller number of key nodes to cause substantial damage to the network than the other three algorithms. Such node combination can have a more important role in maintaining the network structure. To be more specific, the  attack nodes found in the four methods are listed in Table~\ref{Table-1}.

\begin{table}[h]
\centering
\caption{Numbers of attack nodes by different methods. \label{Table-1} }
\begin{tabular}{|l|l|}
\hline
Attack method & Attack nodes\\
\hline
 NIPA & 1,2,3,25,26,33,34 (7 nodes)  \\
 OAS  & 1,2,3,7,10,27,32,33,34 (9 nodes) \\
 HDF  & 1,2,3,4,6,9,14,24,32,33,34 (11 nodes)   \\
 HBF  & 1,2,3,6,9,14,20,32,33,34 (10 nodes)   \\
\hline
\end{tabular}
\end{table}

With seven nodes under attack, we can see that the whole network fragments
and loses a vast majority of its connectivity. In this case, the value of $\kappa $ is 1.9286 (i.e.,$<2$), which deems to lead to network disintegration. In real-world applications, the removal of a node can be costly, thus an attack strategy on an undesired harmful  network should focus on disintegrating the network with fewer attack nodes. On the other hand, the protection of key networks requires to focus on such key nodes or combination of nodes.
This simple experiment shows that our algorithm can be more effective for disrupting
a network, and can also be useful to identify the crucial nodes of a network.

\subsection{Experiments on three different networks}

\subsubsection{Three types of networks}

As mentioned earlier, the BA network is a scale-free network with a special characteristic. That is, most nodes in such networks are connected to only a few nodes, and very few nodes are connected to a very large number of nodes, which means that
some nodes can be potentially more critical than others. Consequently, such scale-free networks can be highly resilient to failure; however, they can be very vulnerable to collaborative attacks, especially such attacks target at certain critical nodes.

A special class of random graphs is the so-called small-world network, which was investigated in detail by Duncan Watts and Steven Strogatz in 1998 \citep{Watts1998Collectivedynamics}. With a probabilistic reconnection of each edge
(with probability $p$), such small-probabilistic, random connections can reduce the average distance between nodes on the network, while maintaining the main structure of the original networks. However, the additional random reconnections can lead to more complex clustering characteristics with a possible phase transition at certain probability, which gives rise to the main characteristics of small-world networks. For example, the ER network model can be used to describe some characteristics of communications and biological networks. In an ER network, the numbers of nodes and connected edges are
fixed, but the connection between each pair of nodes can be random.

Now we will use different network models to further validate our proposed approach.

\subsubsection{Experimental settings}

For a BA generation network, the mean degree ($m$) is set to 3. In addition, the initial mean degree (${m_0}$) of a random network is set to 3, and $p=0.8$ is used. Furthermore, the removal probability $p$  and the mean degree $m$ of a WS small world network are 0.5 and 4, respectively. For an ER network, $p$ is set to 0.02 for a total number of $N=300$ nodes and $p=0.01$ for $N=500$.

In order to compare different methods fairly, the key parameter values are chosen based on our parametric studies and the recommendation from the literature, and these parameter values are summarized in Table~\ref{Table-2}. In the ER experiment, the maximum number of iterations $T=800$ is used for $N=500$ nodes. We have used the population size, which is the same as the network size. This is because we  want to compare our algorithm with other algorithms under the same parameter settings. Obviously, other population sizes can also be used.

In order to ensure that results can be independent of any initial configurations, each algorithm has been executed 10 times. Then, the extent of damage to the network by different methods with different attack strength has been analyzed.
\begin{table}[h]
\centering
\caption{Parameter setting.  \label{Table-2} }
\begin{tabular}{|l|l|}
\hline
\textbf{parameter } & \textbf{value}\\
\hline
 $\alpha$        & 0.3  \\
 Tabu list       & 10   \\
 Network size & $N=300$, $N=500$ \\
 Population size & $n_p=300$ (for $N=300$),  $n_p=500$ (for $N=500$)  \\
 Max-iterations & $T=300$ (for $N=300$),  $T=500$ (for $N=500$)  \\
 BA network & $m=3, m_0=3, p=0.8$    \\
 WS network & $p=0.5, m=4$        \\
 ER network & $p=0.02$ (for $N=300$),  $p=0.01$ (for $N=500$) \\
\hline
\end{tabular}
\end{table}

\subsubsection{Experimental results}

\begin{figure}[ht]
\centering
\includegraphics[width=10cm]{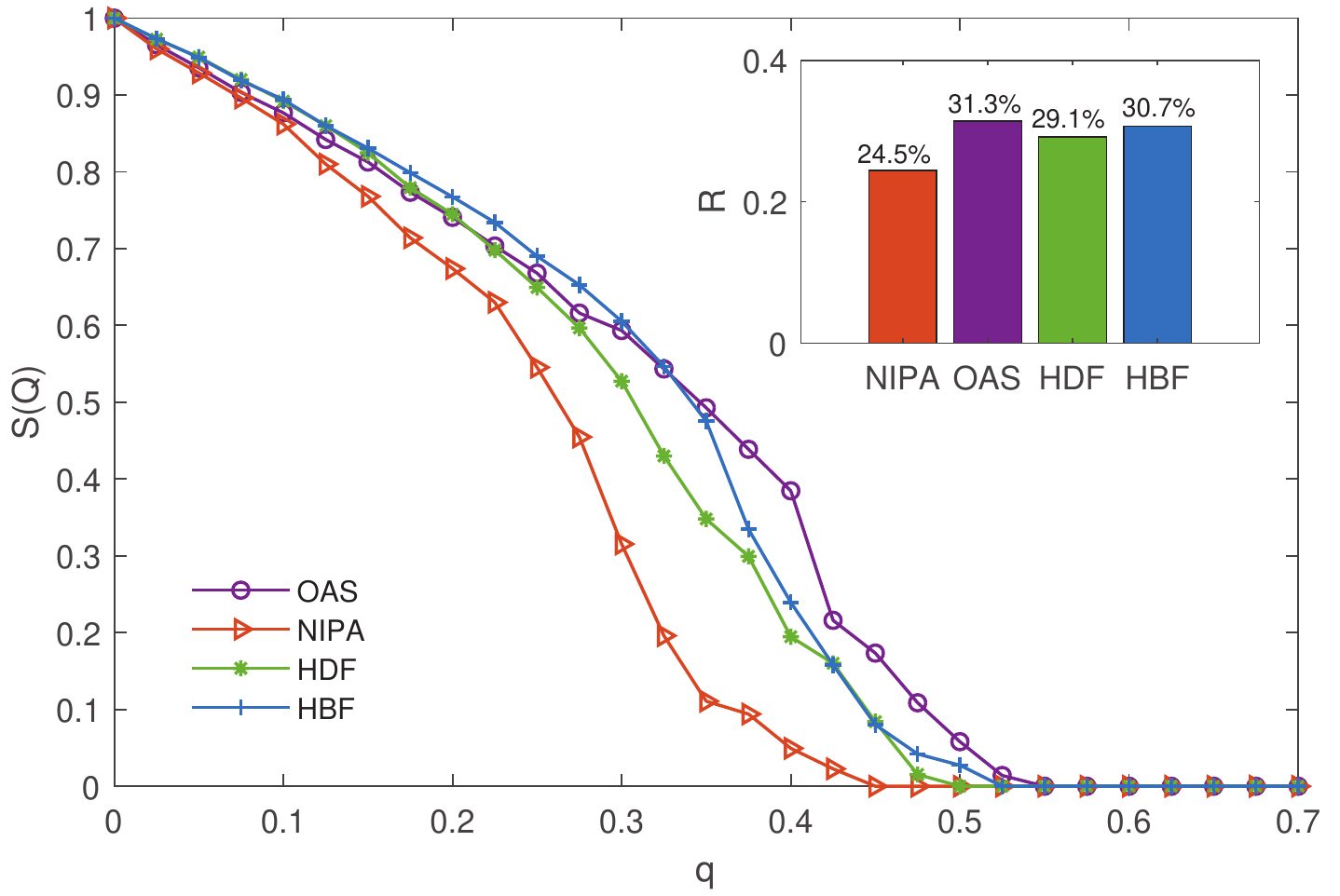}
\caption{Comparison for WS small-world networks with $N = 300$, $p = 0.5$, and $m = 4$. \label{Fig-WS-300} }
\end{figure}

\begin{figure}[ht]
\centering
\includegraphics[width=10cm]{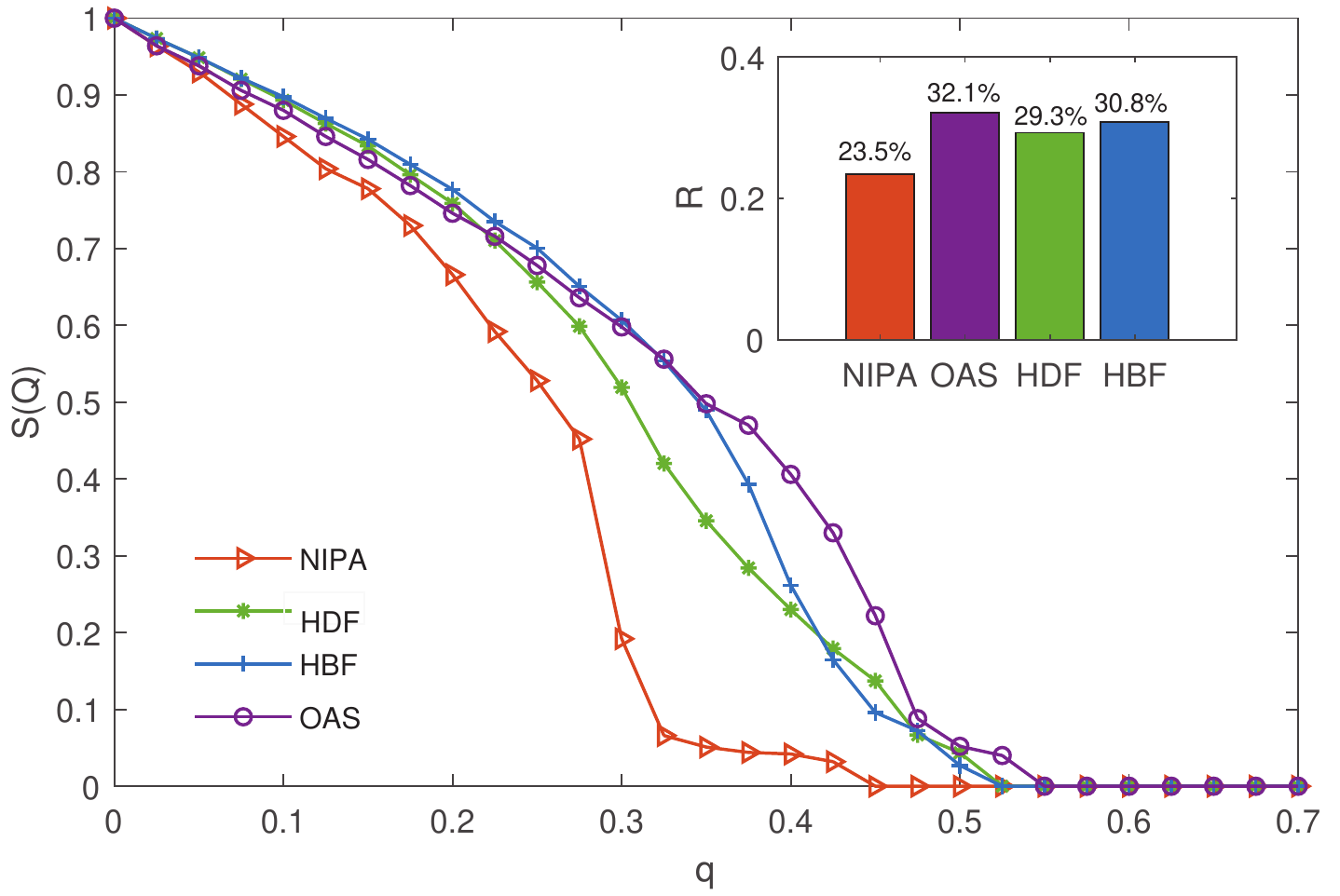}
\caption{Comparison for WS small-world networks with $N = 500$, $p = 0.5$, and $m = 4$. \label{Fig-WS-500} }
\end{figure}

Using the above experimental settings, the results of our numerical experiments are summarized here.
For the WS networks, Fig.~\ref{Fig-WS-300} and Fig.~\ref{Fig-WS-500} show
the comparison of different methods. As we can see from both figures, the proposed algorithm has the most competitive performance in comparison with OAS, HDF and HBF. The results show that the remaining largest cluster after attack by the NIPA is the smallest, which means that the NIPA can incur the most damage to the networks, in comparison with other three methods. From the figures, we can see that the NIPA method has a critical attack intensity $q_c=0.45$ (or $45\%$ of the nodes), while other three algorithms have $q_c$ greater than $50\%$. This means that the NIPA can use a fewer nodes to cause the same damage to the network. In addition, comparing the $R$ values of the four algorithms, it can be observed that the NIPA has a  smaller $R$ than those of other three algorithms.

\begin{figure}[ht]
\centering
\includegraphics[width=10cm]{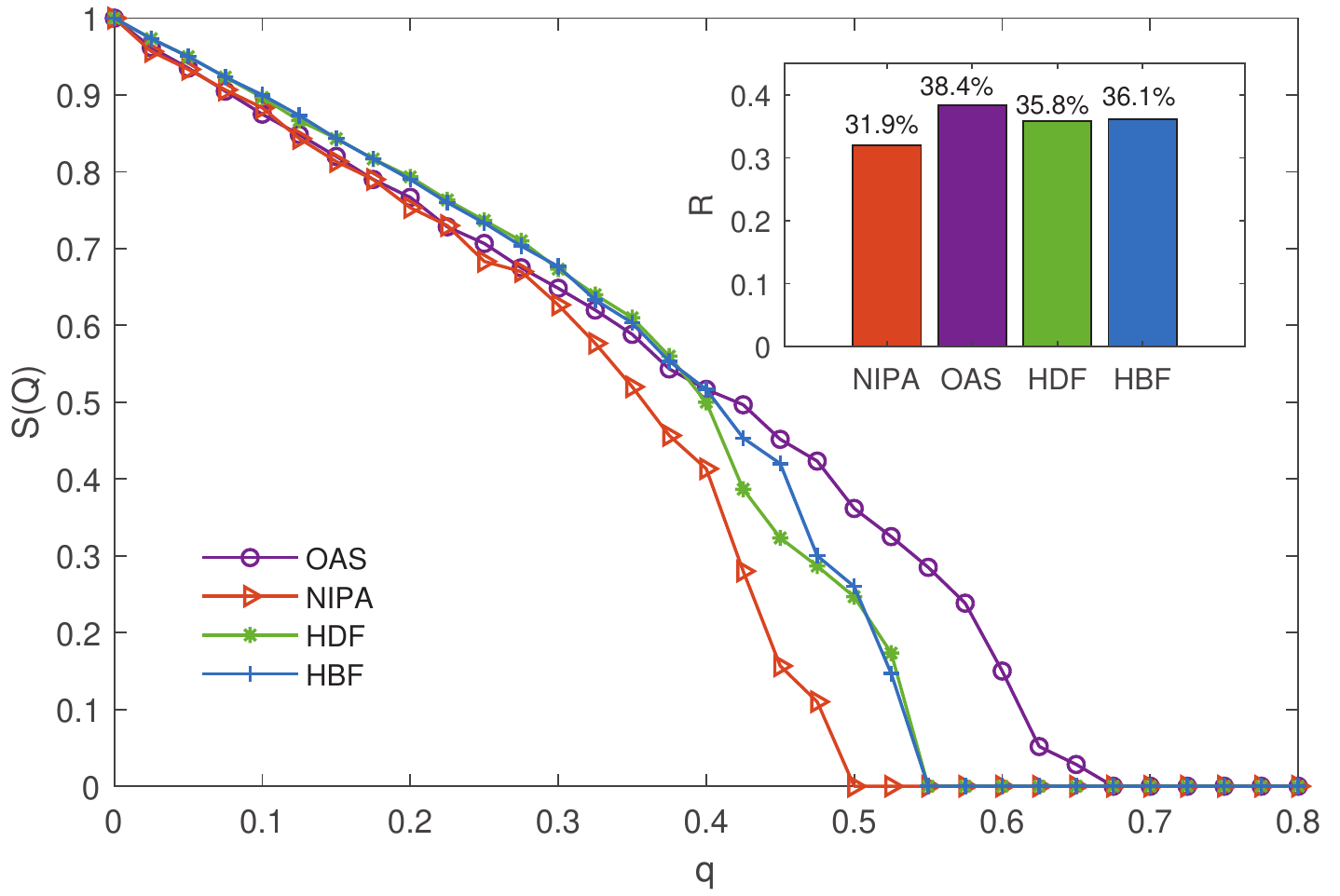}
\caption{Comparison for ER networks with $N = 300$, $p = 0.02$, and $m = 2$.
\label{Fig-ER-300}}
\end{figure}

\begin{figure}[ht]
\centering
\includegraphics[width=10cm]{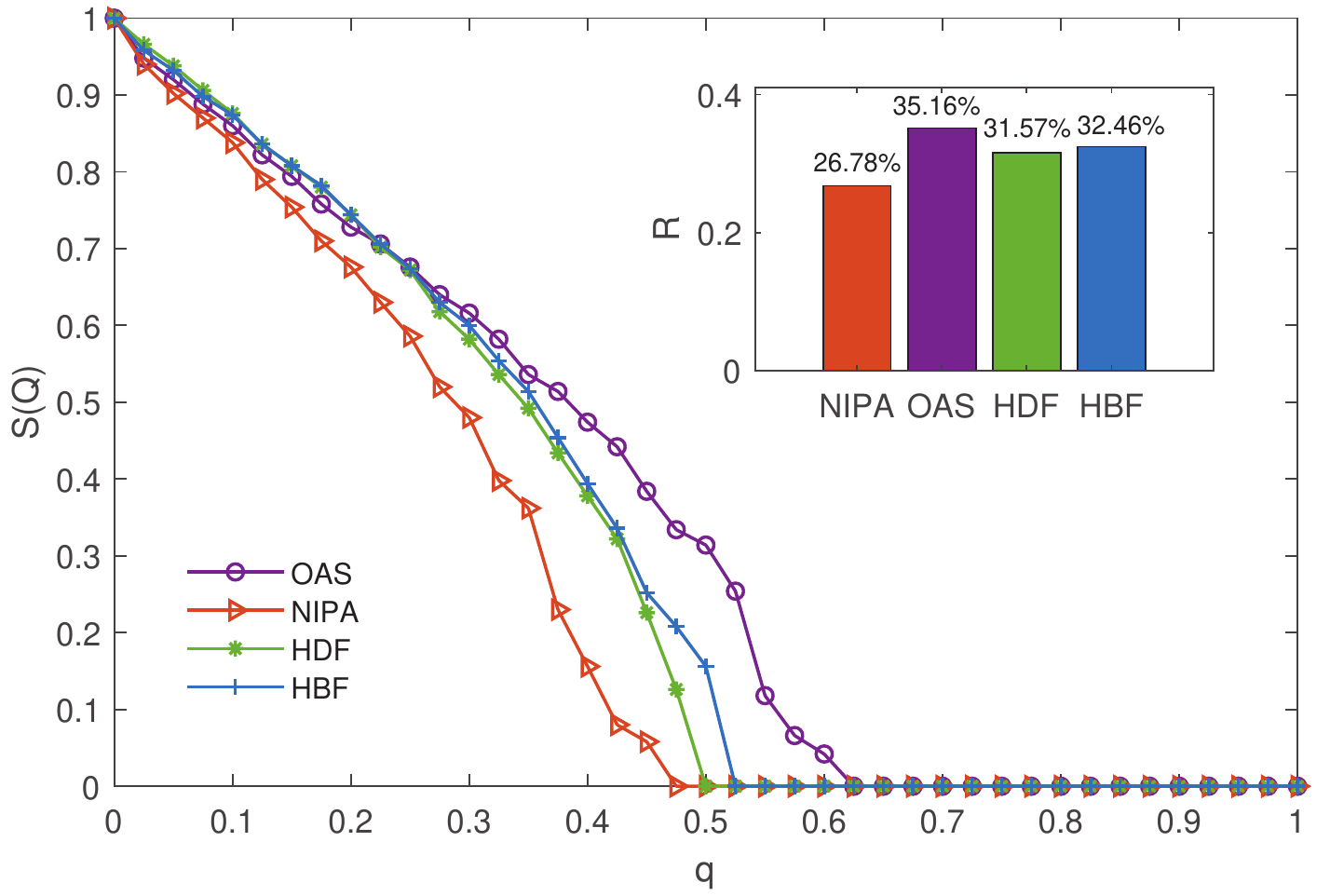}
\caption{Comparison for ER networks with $N = 500$, $p = 0.01$, and $m = 2$.
\label{Fig-ER-500}}
\end{figure}

For the ER networks, we have used the network size of 300 and 500 nodes for two experiments. The results and their comparison are shown in Fig.~\ref{Fig-ER-300} and Fig.~\ref{Fig-ER-500}. Again from these two figures, we can see that the NIPA is more destructive than the other three attack strategies, which becomes evident from the lowest $S(Q)$ and $R$ values. In other words, the NIPA can disintegrate the networks with the least number of attack nodes.

\begin{figure}[ht]
\centering
\includegraphics[width=10cm]{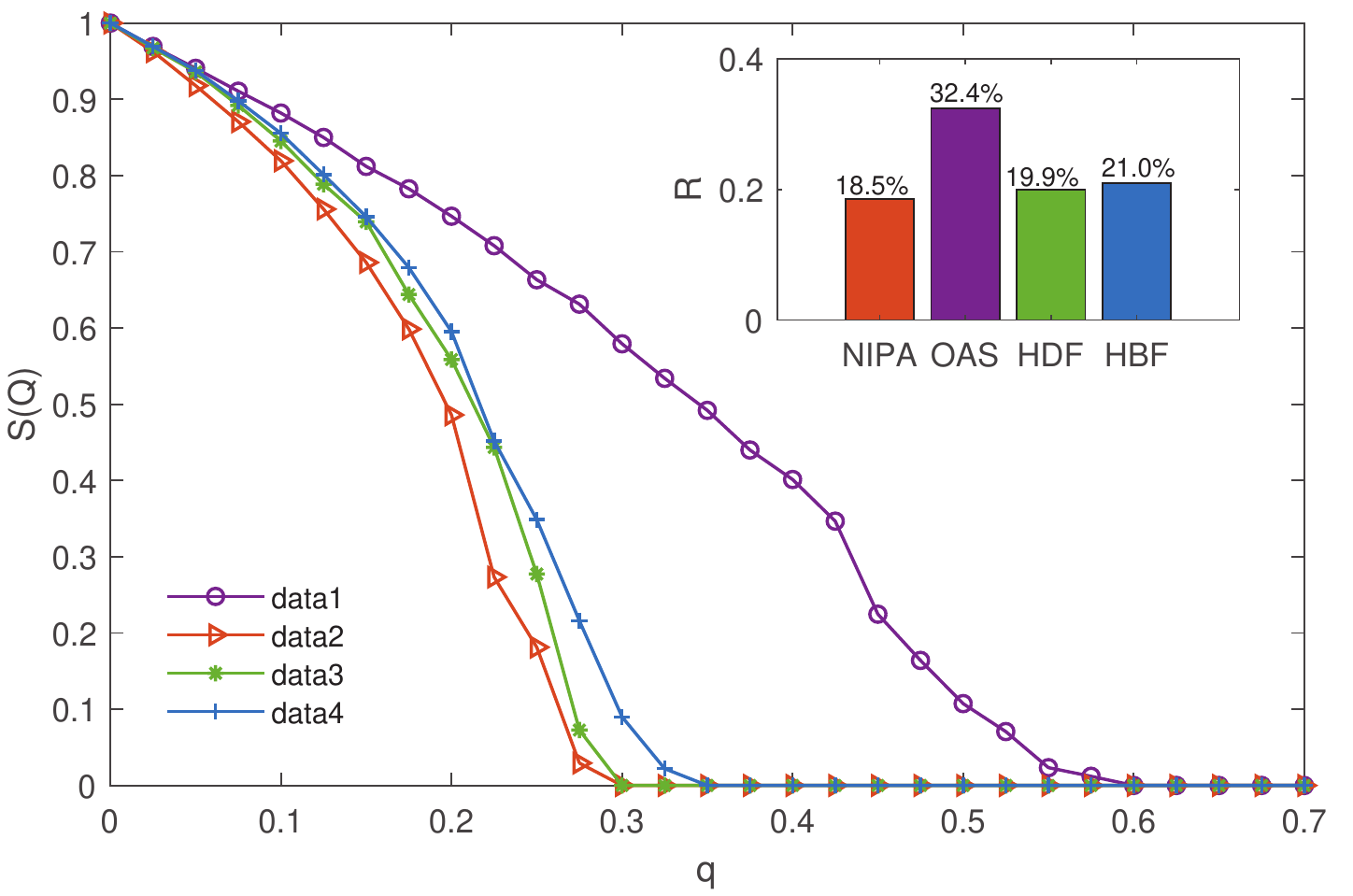}
\caption{Comparison for BA networks with $N = 300$, $p = 0.8$, $m = 3$, and ${n_0} = 3$. \label{Fig-BA-300} }
\end{figure}

\begin{figure}[ht]
\centering
\includegraphics[width=10cm]{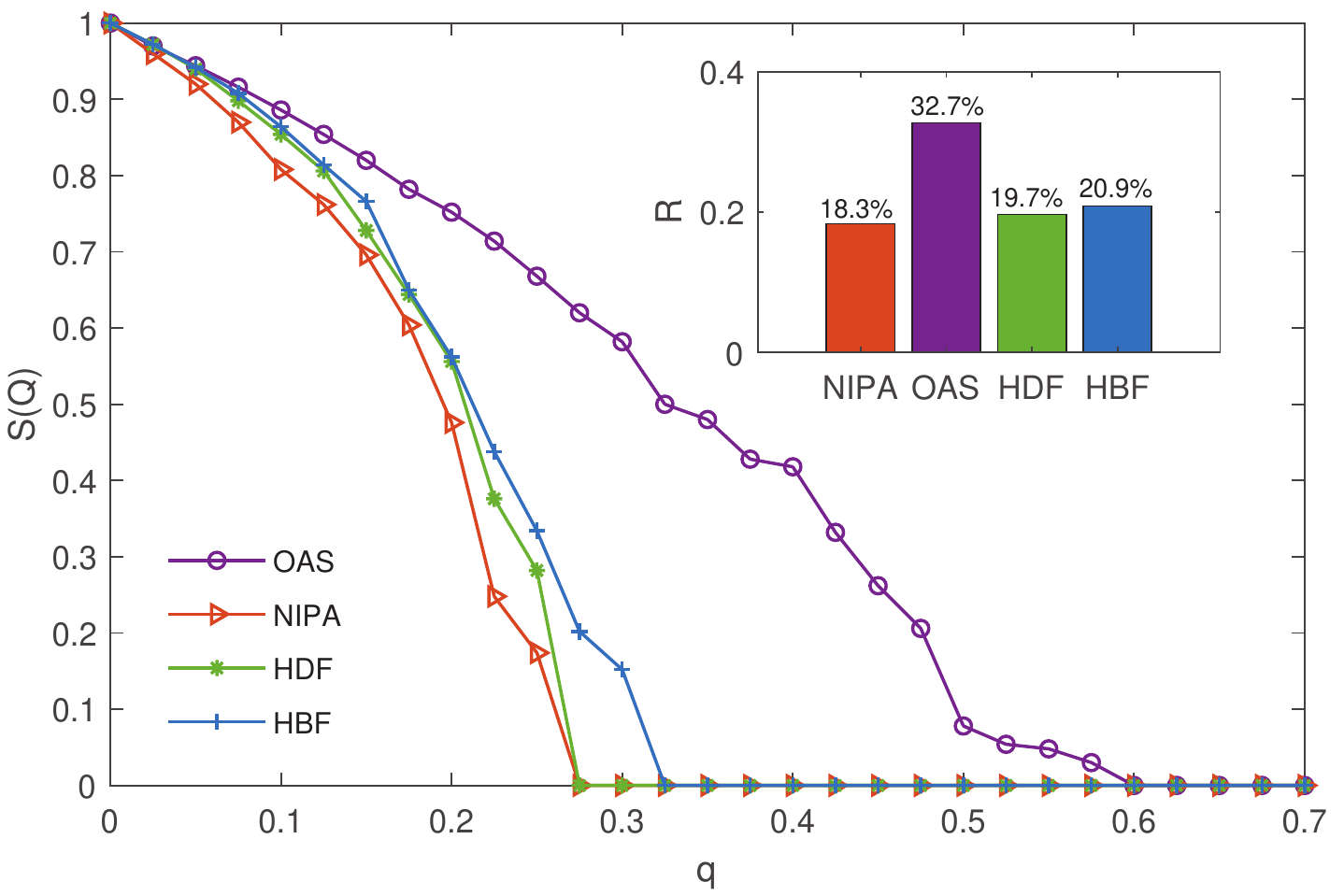}
\caption{Comparison for BA networks with $N = 500$, $p = 0.8$, $m = 3$, and ${n_0} = 3$.
\label{Fig-BA-500} }
\end{figure}

For BA networks, the results and their comparison for four different methods are summarized in Fig.~\ref{Fig-BA-300} and Fig.~\ref{Fig-BA-500}.
The $R$ values of the NIPA on BA networks are the lowest.
In addition, for the same attack intensity, the $S(Q)$ values of our NIPA are lower than those of other methods. Furthermore, when $S(Q)$ first approaches to zero, the value $q$ corresponds to a critical fraction $q_c$, and the $q_c$ value for our proposed NIPA is also the lowest or the same as that obtained by the HDF. One main reason is that BA networks are vulnerable to
collaborative attacks, and thus their weak nodes can be found easily.

{\color{black}The above numerical experiments and comparison studies show that our proposed NIPA can find
a smaller set of optimal key nodes to disintegrate the whole network. In the above figures, the critical change occurs at the point where the curve intersects the horizontal axis, which signifies that the attack causes the collapse of the network. Comparing the convergence curves of the four methods, it can be seen that the NIPA has a much smaller value of $S(Q)$, which indicates that the NIPA is more destructive to the network than other three methods. More importantly, some weakly connected nodes, which are often ignored in other methods, can play an important role because they appear in the optimal set of key nodes. This also shows that our proposed approach can use some vulnerable nodes that are weakly connected so as to increase the overall attack effectiveness, even with a smaller set of key nodes. Conversely, when we aim to protect a crucial network, the key nodes in combination with some weakly connected nodes should be considered for protection. Existing strategies for protection should be revised to focus on the combination of nodes, rather than isolated key nodes.
}

\section{Discussions} \label{S:4}

{\color{black}
Based on the results of the above numerical experiments and a preliminary parametric study, we can now discuss the influence of parameters  of the algorithm, and the computational complexity of our approach. We will also discuss the strength and weakness of our approach so as to inspire further research.

\subsection{Parametric studies}

In our NIPA, a reservation mechanism is used and has been implemented as a percentage parameter $\alpha$. That is, among all target attack nodes,  $\alpha *Q$ nodes remain unchanged in the update mechanism, while the other $(1 - \alpha )*Q$ nodes are updated. Different networks may have different structure and connection characteristics, so we have used three kinds of networks, separately. Five different network instances are selected for each type of networks, and the values of different parameters have been experimented.

The value of $\alpha$ can range from 0.1 to 0.9 with an increment of $0.1$, and $q_c$ is the critical fraction of nodes when the network completely collapses. The results of the experiments are shown in Fig.~\ref{Fig-SW-alpha} where it indicates that, for SW small-world network, the attack effect is the best  when $\alpha=0.6$.

\begin{figure*}[htbp]
\label{fig_bug}
\centering
\subfigure[]{\includegraphics[width=5cm]{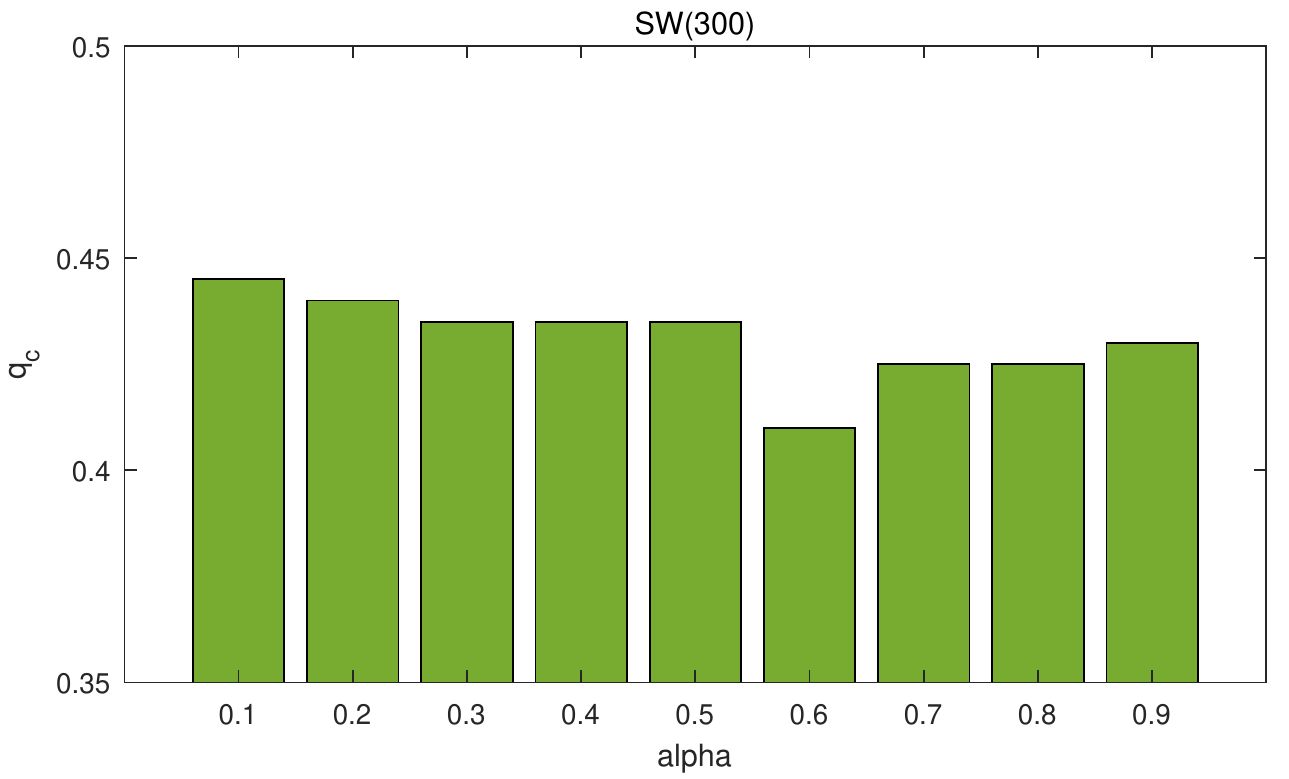}{\label{Fig-SW-alpha}}}
\subfigure[]{\includegraphics[width=5cm]{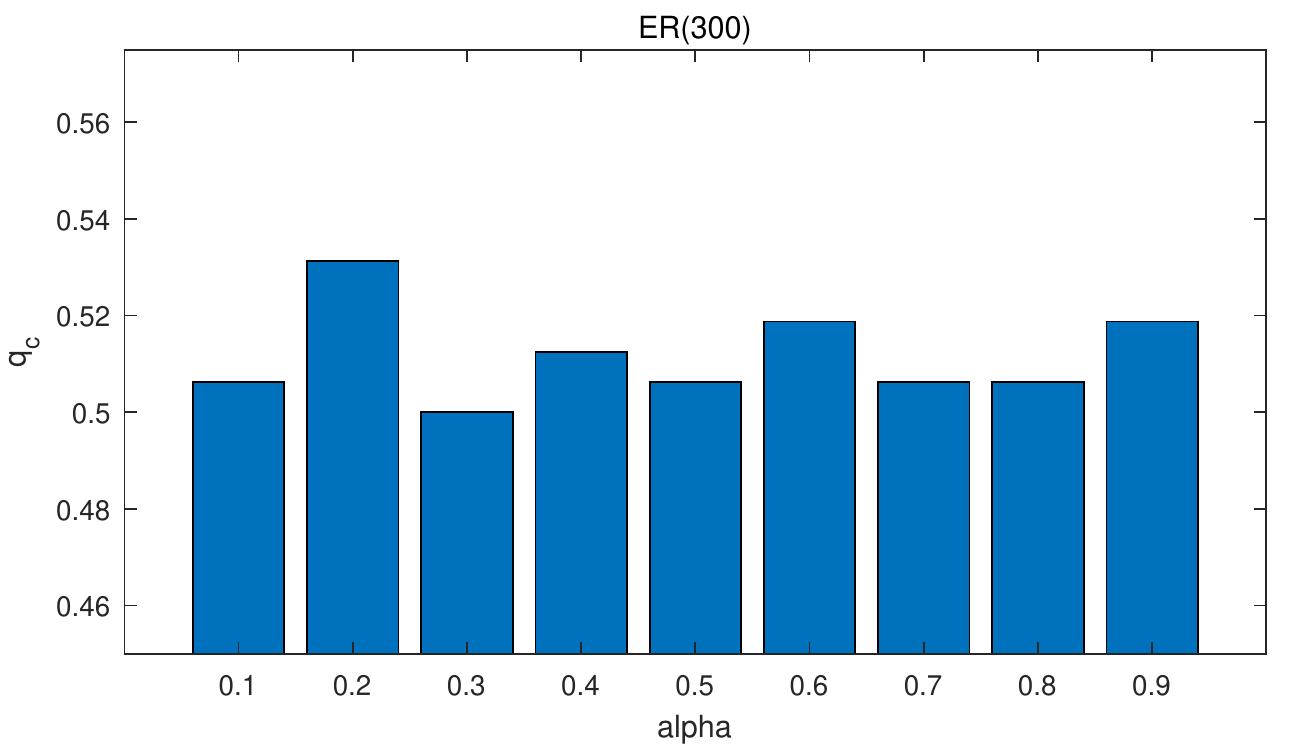}{\label{Fig-ER-alpha}}}
\subfigure[]{\includegraphics[width=5cm]{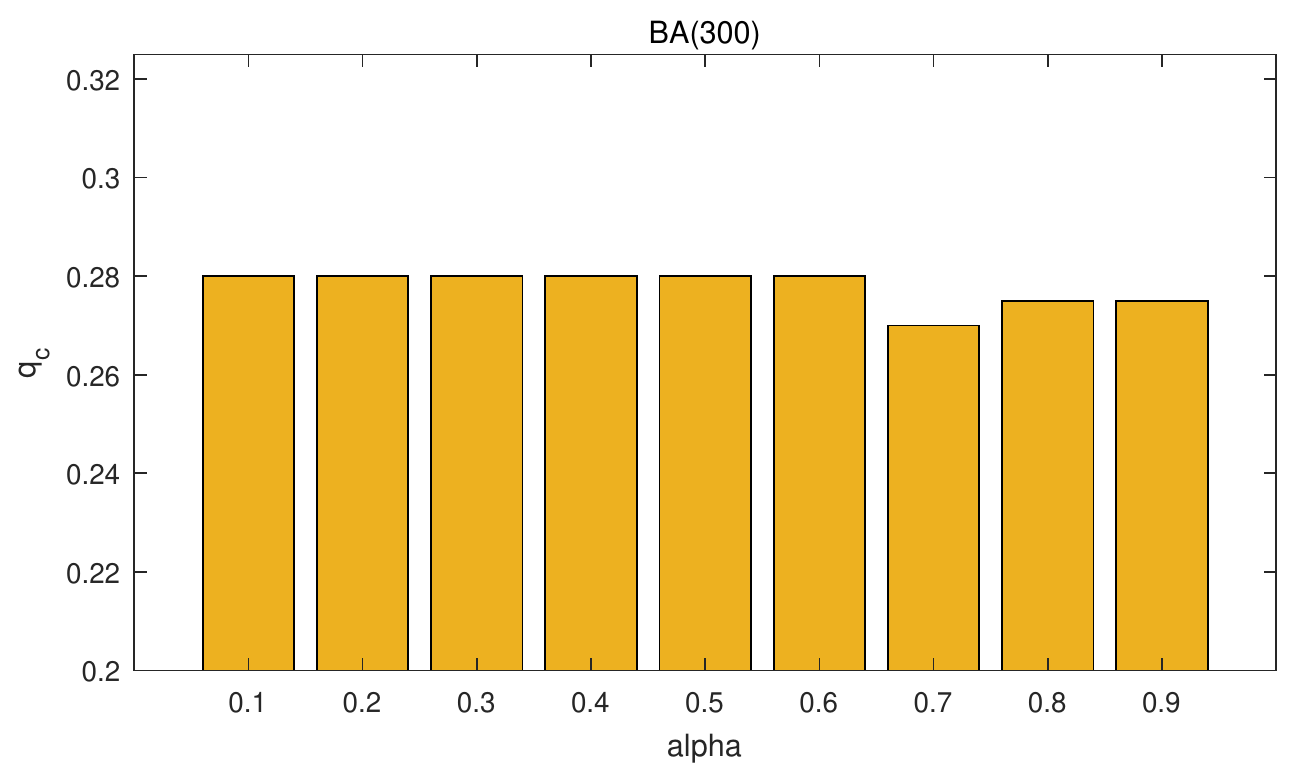}{\label{Fig-BA-alpha}}}
\caption{The experimental results of SW, ER and BA networks with different values of $\alpha$: (a) SW network, (b) ER network, (c) BA network. }
\label{Fig-alpha}
\end{figure*}

For ER networks, the results shown in Fig.~\ref{Fig-ER-alpha} implies that the performance of NIPA is sensitive to the value of $\alpha$. When $\alpha  = 0.3$, the set of attack nodes is smaller than others.
The experimental results are about the same but at the second place when the value of $\alpha$ is 0.1, 0.5, 0.7 and 0.8. In addition, Fig.~\ref{Fig-BA-alpha} shows that a BA network is insensitive to the value of $\alpha$, though the critical fraction $q_c$ of network disintegration is the smallest when $\alpha=0.7$.

From Fig.~\ref{Fig-alpha}, we can see that the value of $\alpha$ does have some influence on the performance of the algorithm. It is worth pointing out that $q_c$ is the percentage of attack nodes in the total number of nodes,  thus for large-scale networks, a small drop as seen in Fig.~\ref{Fig-alpha} will lead to a significant reduction in the number of attack nodes. Consequently, for protection of crucial networks, a small fraction of nodes should be protected with the highest priority.

On the other hand, the population size is also an important parameter. To see if the performance of our  algorithm can largely depend on the population size, different population sizes have been used in the following experiments. For a given network size of 300, five network instances have been randomly selected for each type of network. Population sizes are varied as 100, 200, 300, 400, and 500, while keeping other parameters unchanged. The results are summarized in Fig.~\ref{Fig-pop}.

\begin{figure*}[htbp]
\label{fig_bug}
\centering
\subfigure[]{\includegraphics[width=5cm]{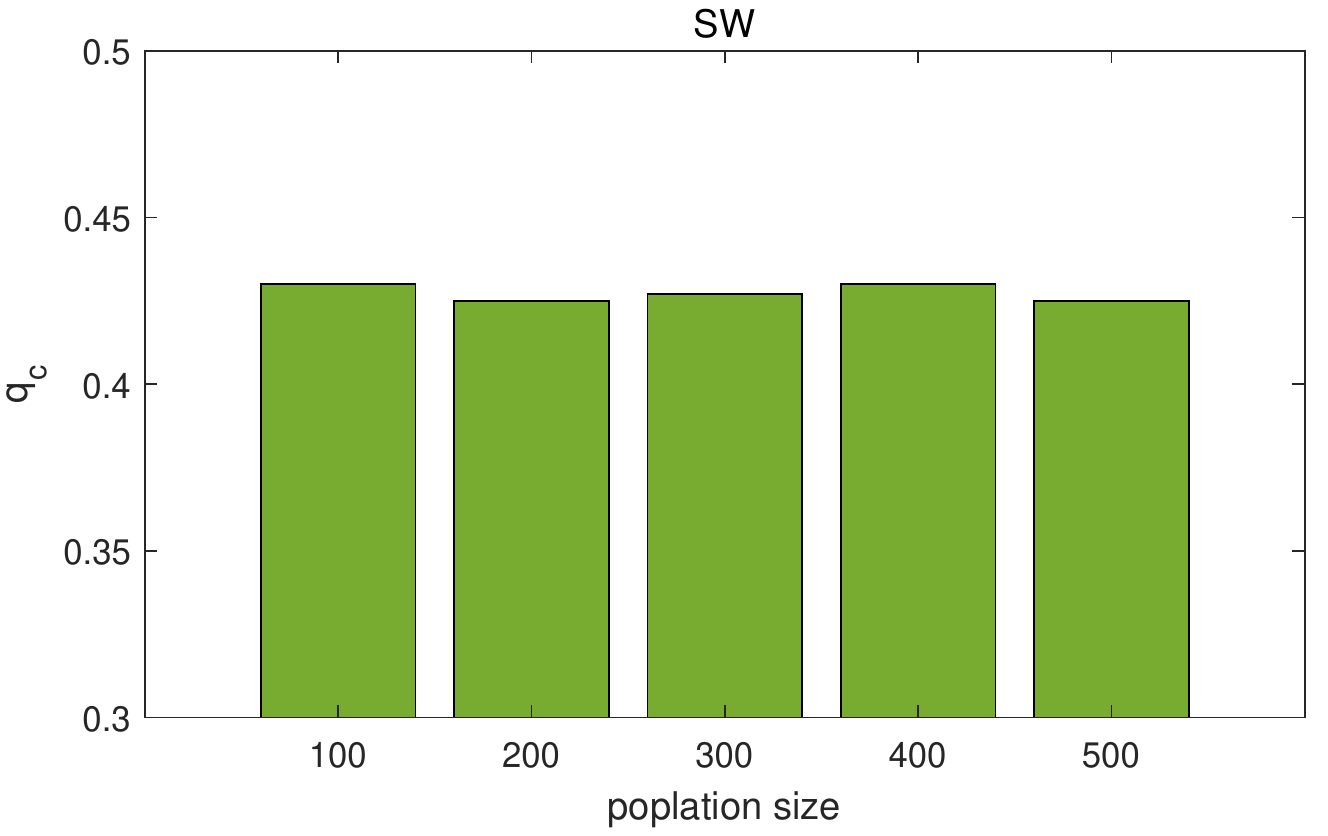}{\label{Fig-SW-pop}}}
\subfigure[]{\includegraphics[width=5cm]{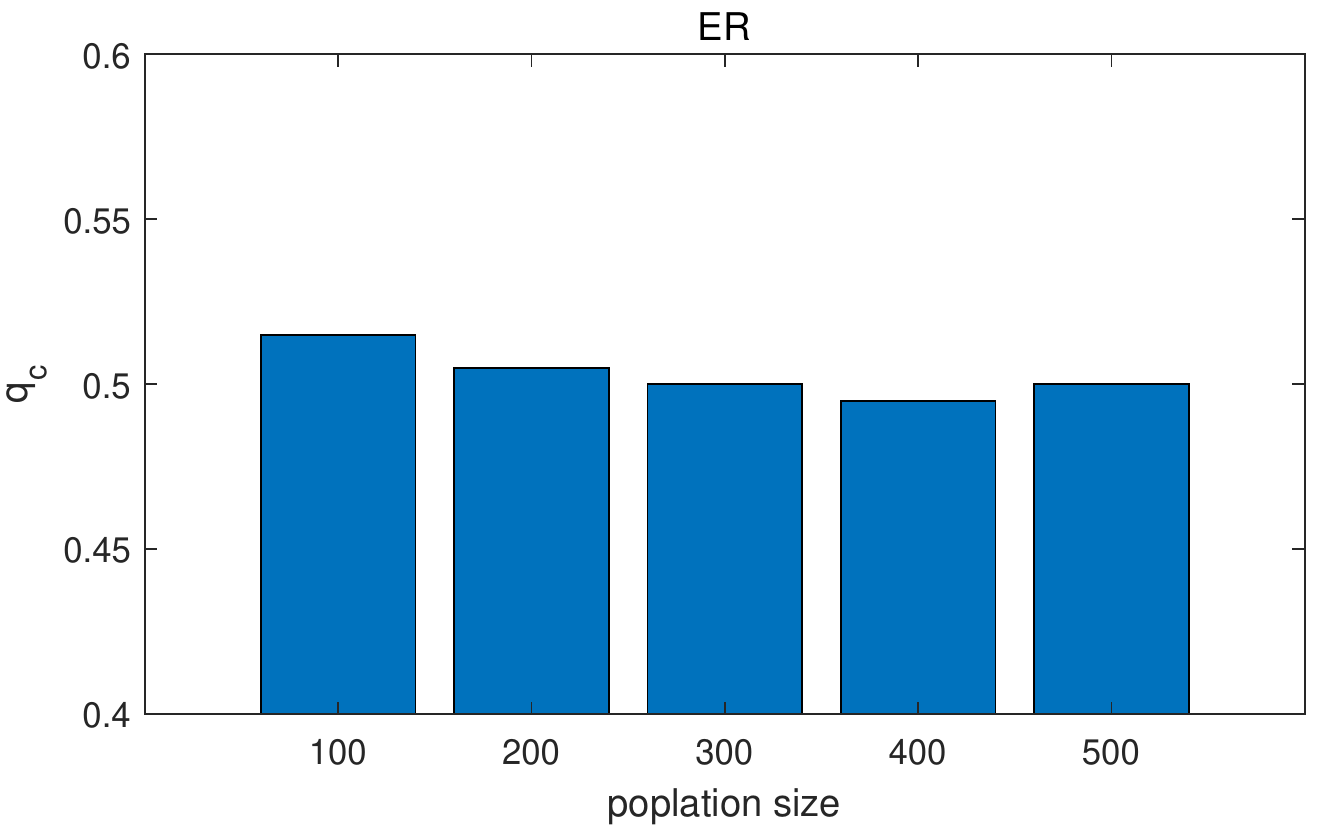}{\label{Fig-ER-pop}}}
\subfigure[]{\includegraphics[width=5cm]{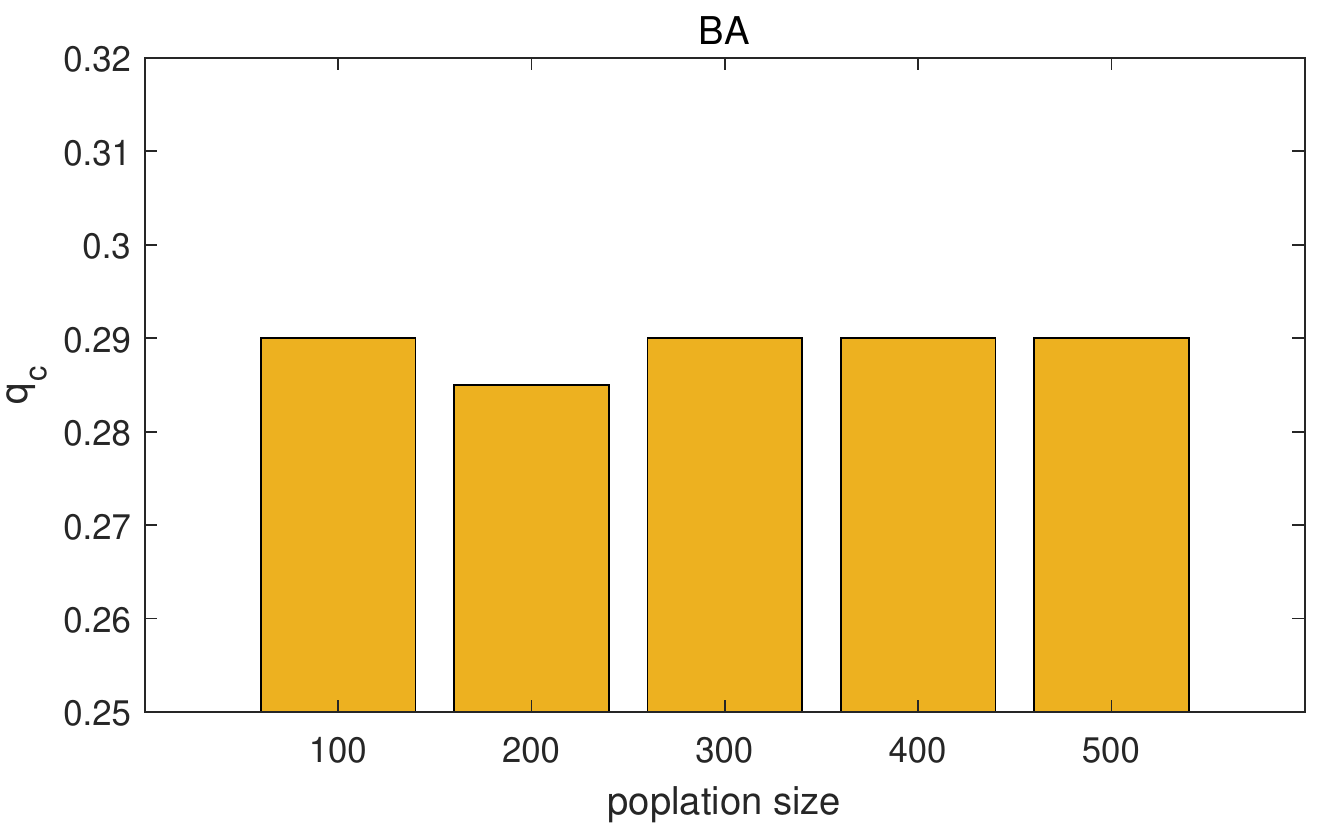}{\label{Fig-BA-pop}}}
\caption{The experimental results of SW, ER and BA networks with different population sizes: (a) SW network,  (b) ER network, (c)BA network. }
\label{Fig-pop}
\end{figure*}

As seen in Fig.~\ref{Fig-pop}, when the population size increases, the value of $q_c$ does not change much, which means that $q_c$ is relatively insensitive to the population size. For SW and BA networks, the accuracy of the algorithm is slightly improved when the population size is 200. For ER networks,
$q_c$ decreases slightly as the population size increases. When the population size is 500, the effect of the algorithm does not continue to improve. Therefore, we can conclude that our proposed NIPA is not particularly dependent on the population size. In addition, for a fixed number of maximum iterations,  only a small size of population is needed, which also means that the number of fitness function evaluations (EFS) is small. This is a strong indication that our proposed algorithm is efficient.

\subsection{Algorithmic complexity}

Now let us estimate the complexity of our algorithm.  The first step in the iterative process is to calculate the attack probability. Since we only calculate the probability of the target attack nodes once at each iteration, the complexity is $O(Q*\left\langle k \right\rangle )$ where $Q(q*N)$ is the attack intensity and $\left\langle k \right\rangle $ is the average degree of nodes in the network.
Then the $Q$ values of probabilities are sorted, and the fast sorting method has been used with the complexity of $O(Q\log Q)$.

The algorithmic complexity for the update steps is relatively low. The complexity of the function evaluations is highly dependent on the complexity of the largest connected cluster calculations. The depth first search (DES) has been used for calculating the largest connected cluster and its complexity is $O({N^2})$. Thus, the algorithmic complexity at each iteration is $O(Q\left\langle k \right\rangle  + Q\log Q + {N^2})$, and the complexity of a single individual in an iteration of the algorithm is at most $O({N^2})$. So the overall complexity of our proposed approach is $O(N^2 t_{\max})$ where $t_{\max}$ is the maximum number of iterations.

\subsection{Strength and weakness of NIPA }

The main contribution of this paper is for the first time to use a probabilistic approach using both neighborhood information and interactions of nodes. We have shown in the above experiments that the proposed approach is efficient. Here, we discuss the strength and weakness of our present work.

One strength of our NIPA is that we use both local nodal information such as degrees and two-hop information such as interactions and neighborhood. This is combined with a probabilistic heuristic algorithm, in contrast with the restricted greedy search used by other methods. Another strength of the proposed approach is that it uses a new centrality-based measure, namely importance measure, thus the present approach emphasizes on the combinations of nodes, rather than simple collection of high-degree nodes. Consequently, weakly connected nodes can become important and such weakly connected nodes are ignored by other methods. As a result, our method is more efficient because a smaller set of key nodes are needed to achieve an optimal attack strategy as shown in our numerical experiments.

However, a slight weakness of the proposed NIPA is that it uses a slightly higher computational costs compared with those of HDF and HBF. This is due to the search of node combinations. Obviously, the effectiveness of the proposed approach and the smaller sets of key nodes it finds far outweigh its weakness.

\section{Conclusions and Further Research Directions} \label{S:5}

The integrity of complex networks is an important issue, and it plays an important role in
many applications concerning networks such as transportation, power transmission, telecommunications, engineering systems and others.  In this paper, we have proposed a probabilistic algorithm based on neighborhood information so as to evaluate different attack strategies and thus ways of protecting crucial networks. Our proposed NIPA uses a new importance measure in combination with a reservation mechanism, and its emphasis is on the node combinations, neighborhood interactions of nodes and two-hope node information, rather than on the centrality-based degrees. We have also carried out a series of numerical experiments using well-known network benchmarks and BA, WS, ER networks. Our simulation results and comparison have shown that the NIPA is more destructive than the other three strategies, given the same attack intensity.

Though the preliminary results are promising, there are are still some issues need to be addressed, which can evaluate the proposed method more thoroughly and also potentially improve its performance even further. These can form the topics for further research.

\begin{itemize}

\item The optimal attack strategy found by algorithms, including the new NIPA, may largely depend on the measures used. Most existing methods use centrality-based measures as a key indicator for network properties such as degrees, closeness, and betweenness. We have used the importance measure in this paper. It will be useful to evaluate and compare different measures for the same networks and same set of algorithms so as to gain insight into different measures and their possible influence on their corresponding algorithm performance.

\item Compared with the methods based on centrality measures, heuristic and probabilistic algorithms tend to have higher computational costs. Thus, any improvement or speed-up is desirable, and this can be achieved by parallelization and cloud computing. In addition, more efficient swarm intelligence-based algorithms such as particle swarm optimization and firefly algorithm can be explored for this purpose.

\item Parameter settings can be important to the performance of an algorithms. The current study has used a basic parametric study. It would be useful to carry out parameter settings more systematically, and this can be combined with sampling methods such as Latin hypercubes and Monte Carlo methods. This will gain a deeper understanding about the sensitivity and robustness of the proposed algorithm and other algorithms.

\item Evaluations and applications of the proposed method and other methods can focus on a diverse range of networks with different properties so as to identify the most effective ways of protecting important networks such as telecommunication networks and dismantling undesired networks such as epidemic diseases networks and terrorist networks.

\item Real networks are very diverse, highly complex and large-scale. It is highly needed to evaluate existing methods using real-world, large-scale networks, and then develop more effective methods
    for studying network integrity and robustness.
\end{itemize}

The above challenges require a multidisciplinary effort by researchers from different research communities  to work on important problems, combining expertise from network theory, computer science, expert systems, artificial intelligence, swarm intelligence, probability theory and system engineering.
It is our hope that this work can inspire more research in the above topics.

}

\section*{Acknowledgement}
This work has been supported by the National Natural Science Foundation of China (Grant Nos.61877046 and 61877047), Shaanxi Provincial Natural Science Foundation of China (Grant No.2017JM1001), the Fundamental Research Funds for the Central Universities, and the Innovation Fund of Xidian University.

{\color{black}
The authors would like to thank the anonymous reviewers for their very constructive comments and suggestions, which have improved the manuscript significantly.}

\section*{Declarations of Interest}
None.

\section*{Highlights}
\begin{itemize}
  \item A new heuristic probabilistic algorithm is proposed
        for evaluating network integrity and robustness.
  \item A novel centrality measure, namely importance measure(IM), is defined and used.
  \item The effect of node combinations is evaluated.
  \item Simulations show that our proposed approach is more effective than other three methods.
\end{itemize}



\bibliographystyle{model5-names}
\small
\bibliography{Qian_refs}

\end{document}